%% file: main.tex
\documentclass[acmlarge]{acmart}

\AtBeginDocument{\providecommand\BibTeX{{\normalfont B\kern-0.5em{\scshape i\kern-0.25em b}\kern-0.8em\TeX}}}

\usepackage[export]{adjustbox}
\usepackage{tabularx}
\usepackage{makecell}
\usepackage{bm}
\usepackage{amsmath,amsfonts}
\usepackage{algorithmic}
\usepackage{graphicx}
\usepackage{textcomp}
\usepackage{xcolor}
\usepackage{multirow}
\usepackage{hyperref}
\usepackage{booktabs}
\usepackage{hhline}
\usepackage[shortlabels]{enumitem}
\usepackage{subcaption}
\usepackage{soul}
\usepackage{calc}
\usepackage{tikz}

\usepackage{geometry}
\usepackage[utf8]{inputenc}
\usepackage{color, colortbl}

\setcopyright{cc}
\setcctype{by}
\acmJournal{IMWUT}
\acmYear{2026} \acmVolume{10} \acmNumber{4} \acmArticle{229}
\acmMonth{12} \acmDOI{10.1145/3857993}

\newcommand{\best}[1]{\colorbox[HTML]{caebc0}{$#1$}}
\newcommand{\projectname}[1]{EduGage}

\newcommand{\edit}[1]{{\textcolor{black}{#1}}}

\begin{document}

\title[EduGage: Multimodal Engagement Assessment in Video Learning]
{EduGage: \edit{A Multimodal Dataset and Benchmark} for Sensor-Based Momentary Assessment of Engagement in Self-Guided Video Learning}

\author{Zikang Leng}
\authornote{Equal Contribution}
\email{zleng7@gatech.edu}
\orcid{0000-0001-6789-4780}
\affiliation{\institution{School of Interactive Computing, Georgia Institute of Technology}
  \city{Atlanta, Georgia}
  \country{USA}}

\author{Edan Eyal}
\authornotemark[1]
\email{eeyal3@gatech.edu}
\orcid{0009-0000-5072-1357}
\affiliation{\institution{College of Computing, Georgia Institute of Technology}
  \city{Atlanta, Georgia}
  \country{USA}}

\author{Yingtian Shi}
\email{yshi457@gatech.edu}
\orcid{0000-0001-8733-7041}
\affiliation{\institution{School of Interactive Computing, Georgia Institute of Technology}
  \city{Atlanta, Georgia}
  \country{USA}}

\author{Jiaman He}
\email{jiaman.he@student.rmit.edu.au}
\orcid{0009-0007-2817-7675}
\affiliation{
    \institution{School of Computing Technologies, RMIT University}
    \city{Melbourne}
    \country{Australia}
}

\author{Yaqi Liu}
\email{yliu3387@gatech.edu}
\orcid{0009-0009-1997-6592}
\affiliation{\institution{College of Computing, Georgia Institute of Technology}
  \city{Atlanta, Georgia}
  \country{USA}
}

\author{Thomas Plötz}
\email{thomas.ploetz@gatech.edu}
\orcid{0000-0002-1243-7563}
\affiliation{\institution{School of Interactive Computing, Georgia Institute of Technology}
  \city{Atlanta, Georgia}
  \country{USA}}

\renewcommand{\shortauthors}{Zikang Leng, Edan Eyal, Yingtian Shi, Jiaman He, Yaqi Liu, \& Thomas Plötz}

\begin{abstract}
\input{abstract/main}
\end{abstract}

\ccsdesc[500]{Human-centered computing~Ubiquitous and mobile computing}
\ccsdesc[500]{Computing methodologies~Artificial intelligence}

\keywords{ Engagement Prediction, Self-guided Learning, Multimodal Sensing, Wearable, Students}

 \maketitle

\input{sections/intro/main}

\input{sections/background/background}

\input{sections/study/main2}

\input{sections/experiment/main}

\input{sections/discussion/main}

\input{sections/conclusion/main}

\input{sections/ackowledgement/main}

\bibliographystyle{ACM-Reference-Format}
\bibliography{./bibs/ref}

\newpage
 \input{sections/appendix/main}

\end{document}

%% file: abstract/main.tex
\noindent
Engagement, which links to attentional, emotional, and cognitive dimensions, plays an important role in learning. In online and video-based learning environments, learners often need to regulate their own interactions with instructional materials. Measuring and reflecting on engagement can therefore support both learners and adaptive learning systems. In this study, we use wearable and camera-based sensing devices to collect physiological and motion signals, including PPG, ECG, EDA, EEG, IMU, heart rate, temperature, and eye-tracking data, to estimate learner engagement. We conducted a user study with 16 participants in a video-based learning scenario, where participants completed learning tasks and provided repeated in-situ \edit{engagement-probe ratings on a 1--5 scale, reporting how difficult it was to pay attention during the preceding minute}. We \edit{establish a benchmark for} engagement estimation, compare different sensing modalities, and further analyze the feasibility and effectiveness of multimodal modeling for characterizing learner engagement. Across participant-based cross-validation, \edit{the modality-aware reference model} achieves an MAE of \edit{0.80, 83.18\% within-1 accuracy, 70.96\% binary accuracy, and 62.90\% binary Macro-F1}, outperforming sensor-free, statistical, deep temporal, foundation-model, and LLM-based baselines. \edit{Our results suggest that fine-grained engagement estimation is feasible but inherently noisy, and that the preferred sensing configuration depends on the target metric and practical sensing burden.} We release the \projectname{} dataset \edit{as the primary contribution of this work}, including synchronized multimodal sensor signals, probe-aligned \edit{engagement-probe ratings}, video metadata, quizzes, and study materials, to support reproducible research on fine-grained sensor-based engagement modeling in self-guided learning.

%% file: sections/intro/main.tex
\section{Introduction}

Learning is not simply a matter of being exposed to information; its effectiveness greatly depends on how learners engage with that information during the learning process~\cite{krathwohl2002revision}. Engagement is widely understood to involve behavioral, emotional, and cognitive dimensions, each of which shapes how learners participate in and make sense of instructional material~\cite{fredricks2004school,blumenfeld2006motivation}. 
Among these, cognitive engagement is particularly important for meaningful learning because it reflects the extent to which learners invest mental effort, connect ideas, and regulate their understanding~\cite{blumenfeld2006motivation}. Prior work suggests that deeper engagement with learning materials leads to stronger learning outcomes than passive exposure alone~\cite{chi2014icap}, and research on self-regulated learning further links these processes to academic performance~\cite{pintrich1990motivational}.

However, engagement is difficult to measure because these dimensions do not always align and are not directly observable~\cite{henrie2015measuring}. For example, learners may appear behaviorally attentive while mind-wandering, or show little visible response while still processing difficult material~\cite{smallwood2015science}. This makes it challenging to infer engagement from behavior alone, especially when the goal is to capture changes at specific moments of instruction.

\edit{Researchers have therefore explored behavioral and physiological signals as indirect indicators of learners' internal states during instruction~\cite{gao2020n}. Such signals can capture processes related to sustained attention, cognitive effort, and affective arousal, including electrodermal, cardiovascular, and neural measures~\cite{fredricks2004school,blumenfeld2006motivation,calvo2010affect,d2012dynamics,boucsein2012electrodermal,berntson1997heart,pope1995biocybernetic}.}

\edit{Prior work on automatic engagement detection has explored visual, physiological, wearable, and multimodal sensing approaches~\cite{whitehill2014faces,monkaresi2016automated,di2018unobtrusive,gao2020n}. However, these studies often focus on classroom environments, broader engagement dimensions, or coarse labels such as session-level ratings and observer-based annotations~\cite{disalvo2022reading}. It therefore remains unclear whether sensing can support fine-grained, segment-level engagement tracking in self-guided learning contexts, and which sensing configurations offer the best balance between predictive value and practical deployment.}

\edit{This gap is particularly important for college students learning from prerecorded instructional videos in self-guided and asynchronous settings, where continuous instructor feedback is unavailable~\cite{guo2014video,kizilcec2013deconstructing}. In such environments, learners must regulate their own attention, motivation, and understanding as content unfolds~\cite{henrie2015measuring,conrad2021measuring}. Aggregate or session-level engagement scores may obscure when attention declines, which video segments are difficult to follow, or where learners may need support. Fine-grained engagement sensing could therefore support post-hoc content refinement, adaptive learning systems, and learner self-reflection without replacing the role of teachers~\cite{guo2014video,conrad2021measuring}.}

\edit{Wearable and camera-based sensing make this perspective practical because they can collect continuous, low-burden signals during ordinary computer-based video learning. However, practical deployment still requires understanding which signals and models provide useful predictive value without imposing unnecessary sensing burden.}

\edit{In this work, we study fine-grained engagement detection in self-guided video-based learning using wearable sensing. We conduct a user study that captures multimodal signals and repeated, low-burden self-reports aligned to short video segments.} \edit{The resulting data form the \projectname{} dataset, which we release together with study materials and participant-independent evaluation splits.} Using these data, we \edit{establish a benchmark for} estimating engagement over time and comparing different \edit{modeling approaches,} wearable modalities, and device configurations.

Our contributions are as follows:
\begin{itemize}
\item \edit{As our primary contribution,} we collect and release the \projectname{} dataset, a multimodal dataset for fine-grained engagement modeling in self-guided video-based learning, combining wearable \edit{and camera-based sensing streams} with repeated, segment-level self-reports\edit{, quiz measures, study materials, and participant-grouped evaluation splits}. Dataset available at \url{https://doi.org/10.6084/m9.figshare.32145994}.

\item We \edit{establish a participant-independent benchmark for} fine-grained engagement estimation over time, \edit{spanning sensor-free, statistical, deep temporal, foundation-model, LLM-based, and modality-aware approaches.} Code available at \url{https://github.com/ZikangLeng/EduGage}.

\item We systematically compare wearable modalities and device configurations to characterize tradeoffs between predictive performance and deployment feasibility.

\item We provide empirical insights into fine-grained engagement estimation, including the effectiveness of different modeling approaches and the practical limits of wearable sensing.
\end{itemize}

%% file: sections/background/background.tex
\section{Background and Related Work}
\label{sec:related work}

\edit{Our work is positioned at the intersection of learning theory, moment-to-moment engagement measurement, and sensing-based engagement assessment in self-guided video learning. We review engagement in instructional-video settings (Section~\ref{sec:engagement_in_video}), repeated in-situ self-report as a way to capture temporal variation (Section~\ref{sec:temporal engagement}), and prior sensing-based approaches for automatic engagement assessment, including physiological and behavioral correlates of engagement-related processes (Sections~\ref{sec:physiological of engagement} and~\ref{sec:automatic engagement assessment}). Together, this literature motivates our focus on fine-grained engagement estimation using wearable and camera-based sensing under practical deployment constraints.}

\subsection{Engagement in Self-Guided Video Learning}
\label{sec:engagement_in_video}

Student engagement is widely understood as a multidimensional construct comprising behavioral, emotional, and cognitive components~\cite{fredricks2004school,wong2022student}. Behavioral engagement refers to observable participation, emotional engagement captures affective responses, and cognitive engagement reflects learners' mental effort in understanding and mastering material~\cite{kahu2013framing}. Prior work has emphasized cognitive engagement because it is closely tied to deep processing, self-regulation, and knowledge construction~\cite{blumenfeld2006motivation,appleton2006measuring,chi2014icap,pintrich1990motivational}. Thus, engagement is not only visible participation, but also how learners allocate attention and cognitive effort while making sense of instruction.

This perspective is especially important in higher-education settings, where engagement is increasingly understood not as a fixed learner trait, but as a context-dependent process shaped by the interaction between learners, tasks, and learning environments~\cite{kahu2013framing,kahu2018student}. Such a view is particularly relevant for self-guided video-based learning, where learners must regulate their own attention, motivation, and understanding without continuous instructor feedback~\cite{wong2022student}. Instructional videos are now \edit{commonly used in} higher education and professional training, and prior work suggests that their effectiveness depends not only on access to content, but also on how learners cognitively engage with the material as it unfolds over time~\cite{guo2014video,noetel2021video,kuhlmann2024students}. Because video-based learning is inherently sequential, learners may respond differently to different moments, explanations, or transitions within the same lesson, making engagement during video learning especially sensitive to temporal and contextual variation~\cite{guo2014video,kim2014understanding,anders2024associations}.

At the same time, cognitive engagement, particularly its attentional component, is only partially observable from external behavior~\cite{smallwood2015science}. Learners may appear attentive while mind wandering, or may show limited outward activity while still processing difficult material. Because attention can decouple from the task in ways that are not reliably visible~\cite{szpunar2013interpolated,conrad2021measuring}, engagement in self-guided video learning is best understood as a dynamic, partially latent process rather than a fixed state directly inferred from behavior alone.

\subsection{Temporal Engagement and Momentary Self-Report}
\label{sec:temporal engagement}

A growing body of educational research has emphasized that engagement is dynamic, varying over time and across instructional contexts~\cite{shernoff2014student,shernof2017student,martin2020factors}. Experience-sampling studies show moment-to-moment variation in attention, interest, and perceived challenge within lessons~\cite{shernoff2014student}, and later work extends this view across students, lessons, activities, and timescales~\cite{shernof2017student,martin2020factors,xie2019examining}. These findings make aggregate or session-level summaries insufficient for understanding how learners respond to specific moments of instruction.

Because such fluctuations unfold during ongoing activity, capturing them requires temporally localized measurement. Experience sampling and ecological momentary assessment provide a basis for repeatedly measuring subjective states in situ while reducing reliance on retrospective recall~\cite{csikszentmihalyi1987validity,shiffman2008ecological}. In educational settings, these approaches have been used to study engagement within lessons and technology-mediated learning activities~\cite{manwaring2017investigating,xie2019examining,de2024measuring}. Because repeated measurement must be balanced against burden and disruption, dense temporal sampling often relies on brief, low-burden probes, including single-item measures~\cite{dejonckheere2022assessing,song2023examining}. Recent work suggests that such measures can provide useful indicators of subjective states when interpreted as pragmatic approximations rather than exhaustive assessments of complex constructs~\cite{dejonckheere2022assessing,song2023examining,de2024measuring}. \edit{Established engagement instruments typically conceptualize engagement as a multidimensional construct and assess multiple behavioral, emotional, and cognitive components~\cite{fredricks2004school,skinner2009motivational,de2024measuring}. Our single-item measure is not intended to reproduce these comprehensive instruments or to measure engagement in its entirety. Instead, it focuses specifically on learners' perceived difficulty sustaining attention during the immediately preceding segment of instruction. We treat this as a narrower, engagement-related attentional experience rather than as engagement itself or its direct inverse. In particular, a learner may remain highly engaged with demanding material while still finding it difficult to sustain attention, and attention may feel easy for reasons that do not necessarily indicate strong cognitive or emotional engagement.}

\edit{The item was intended to capture the subjective effort required to maintain attention rather than the intensity of concentration. A concentration-focused item, such as asking how hard a learner was concentrating, could receive a high rating during both productive engagement and effortful struggle. By contrast, asking about difficulty sustaining attention targets the learner's perceived struggle to remain focused. Prior instructional research provides motivating, though not definitive, evidence for this formulation: instructional formats described as more engaging have been associated with lower reported difficulty paying attention and stronger learning outcomes~\cite{miller2013comparison}. We therefore use attention difficulty as a temporally localized reference signal related to one attentional aspect of engagement, while recognizing that it does not capture the full engagement construct.}

\subsection{Physiological and Behavioral Correlates of Engagement-Related Processes}
\label{sec:physiological of engagement}

Because engagement is partially latent, sensing-based approaches rely on indirect correlates such as autonomic arousal, sustained attention, cognitive effort, mind wandering, and behavioral restlessness~\cite{calvo2010affect,berntson1997heart,boucsein2012electrodermal}. Physiological and behavioral signals linked to these processes are therefore plausible sources for studying moment-to-moment variation during learning.

Electrodermal activity (EDA) is widely used to study arousal-related processes because skin conductance reflects sympathetic nervous system activity and has been linked to attentional and affective responses~\cite{critchley2002electrodermal,boucsein2012electrodermal,di2018unobtrusive,horvers2021detecting}. Cardiovascular signals, including heart rate, heart-rate variability, ECG, and PPG-derived measures, have likewise been associated with workload, arousal, and attentional demand~\cite{berntson1997heart,cinaz2013monitoring,beh2021robust,mach2022assessing}.

Neural and behavioral signals provide complementary perspectives. EEG-based systems have been used to estimate engagement, workload, and attentional state, including in online lecture and multimedia settings~\cite{pope1995biocybernetic,conrad2021measuring,dan2017real,sarailoo2022assessment,pei2025design}. Posture, movement, and gaze behavior can reveal overt correlates of learner interest, attentional allocation, and mind wandering~\cite{mota2003automated,hutt2017gaze,hutt2019automated,mach2022assessing,he2026characterizing,he2025characterising,liu2024calibread,liu2025enhancing}. Together, these modalities motivate multimodal sensing while raising the practical question of which configurations are most informative and deployable.

\subsection{Automatic Engagement Assessment with Sensing}
\label{sec:automatic engagement assessment}

Building on these modality-specific correlates, prior work has explored automatic engagement assessment using vision-based, physiological, wearable, and multimodal sensing~\cite{booth2023engagement,bustos2022wearables}. Early systems often relied on facial expressions and other visible behaviors~\cite{whitehill2014faces,bosch2016detecting,savchenko2022classifying}, while later work incorporated physiological and multimodal signals such as video-derived heart rate, wearable EDA, classroom multimodal sensing, and in-the-wild prediction of emotional, behavioral, and cognitive engagement~\cite{monkaresi2016automated,di2018unobtrusive,gao2020n,disalvo2022reading}. EEG-based systems have also been used to estimate engagement and related cognitive states, though typically with more intrusive instrumentation~\cite{pope1995biocybernetic,apicella2022eeg}. Multimodal learning analytics likewise emphasizes the value of heterogeneous sensor streams, while noting their interpretive and practical challenges~\cite{ochoa2016editorial,blikstein2016multimodal}.

Despite this progress, several limitations remain. Many engagement-sensing systems rely on coarse-grained labels, post-hoc annotations, or session-level judgments, making it difficult to study fine-grained fluctuations during specific moments of instruction~\cite{whitehill2014faces,monkaresi2016automated,disalvo2022reading}. Much of the literature also centers on classrooms or controlled laboratory environments rather than self-guided instructional-video settings~\cite{gao2020n,gao2022individual,disalvo2022reading}. Finally, practical deployment remains challenging: cameras raise privacy concerns, EEG is often impractical for long-term use, and wearable reviews emphasize device heterogeneity, ecological validity, and deployment feasibility~\cite{bustos2022wearables,schroeder2023scoping,mach2022assessing}.

\edit{Taken together, prior work suggests that engagement-related states can be inferred from sensing data, but relatively few studies jointly examine fine-grained temporal measurement, self-guided instructional-video learning, and practical wearable deployment. This gap motivates our study of temporally aligned engagement-probe ratings, multimodal sensing, and reduced-device tradeoffs.}

%% file: sections/study/main2.tex
\section{Study Design
}
\label{sec:study_design}

\edit{To investigate whether sensing can support fine-grained engagement estimation in self-guided video-based learning, we designed a controlled laboratory study to collect temporally aligned multimodal signals and repeated self-reports during self-guided instructional-video learning. This section describes the participants, learning materials, engagement measure, sensing setup, study procedure, and resulting \projectname{} dataset.}

\edit{\textbf{Terminology.} In this paper, we use \emph{study session} to refer to one participant's complete approximately 90-minute visit, whereas a \emph{video-viewing session} refers to the participant watching one approximately 10-minute instructional video. A \emph{video segment} is the approximately one-minute portion preceding a probe and referenced by its attention-difficulty rating. A \emph{probe-aligned window} is the 44-second sensor interval immediately preceding the probe and used as one prediction sample.}

\subsection{Participants}
\label{sec:participants}

We recruited 16 college students from a university population through mailing lists and online postings. Participants were required to be at least 18 years old and to have normal or corrected-to-normal vision. The study was approved by the Georgia Institute of Technology Central Institutional Review Board (IRB2025-134). All participants provided informed consent prior to participation and received compensation for their time. \edit{Participants also consented to the release of de-identified study data for research use.} \edit{Participants received a base compensation of \$30 and could earn an additional \$10 performance-based bonus by achieving a score of at least 80\% on the post-video quizzes. This bonus was used to encourage participants to make a genuine effort on the learning assessments.}

Before the study, participants completed a short questionnaire covering demographic background and prior experience with online learning.
Detailed questionnaires are included in Appendix~\ref{app:demographics}.
Because our target scenario is self-guided video-based learning in higher-education contexts, we also collected information about participants' typical weekly engagement with online learning activities. The resulting cohort included students from a range of academic backgrounds and levels of prior exposure to online learning, providing a relevant population for the study context considered in this work. Detailed participant characteristics are summarized in \autoref{tab:participants}.\edit{Thirteen participants studied computing-related subjects, while the remaining three studied Biology, Materials Science and Engineering, and Aerospace Engineering.}
\input{tables/participants}

\subsection{Learning Task and Study Materials}
\label{sec:learning_task}

The learning task was designed to approximate a self-guided asynchronous video-learning scenario. Participants individually watched instructional videos drawn from the MIT Open Learning Library, which builds on MIT OpenCourseWare, a public large-scale collection of freely available university course materials~\cite{mitocw}. We selected video-based learning as our task setting because it reflects an increasingly common instructional format in which learners must regulate their own attention and understanding without real-time feedback from an instructor~\cite{guo2014video,henrie2015measuring,noetel2021video}.

\edit{To elicit natural variation in engagement over time, we selected videos spanning multiple subject areas and instructional styles, including both lecture-like presentations and more visually structured explanatory content. Each video was approximately 10 minutes long. This duration allowed us to collect multiple probe-aligned ratings and observe within-video fluctuations in attention difficulty while keeping the four-video learning task manageable within the approximately 90-minute study session. Presenting four shorter videos rather than one longer lecture also exposed each participant to a range of topics and instructional styles, reducing the extent to which the collected responses were tied to a single video or subject domain~\cite{anders2024associations,kim2014understanding,kuhlmann2024students}.} 

\edit{To diversify the instructional content beyond the primary academic backgrounds represented in our sample, we curated videos from four relatively specialized domains: X-Rays, Aerospace Engineering, Environmental Science, and Business. Most participants studied computing-related subjects. Selecting relatively specialized topics reduced, but could not eliminate, prior familiarity with the material; one participant studied Aerospace Engineering.} For each topic, we selected two video variants, yielding an A/B version pair per topic. \edit{Participants watched four videos in an order determined by a four-sequence Williams design~\cite{williams1949experimental}. The sample of 16 allowed equal allocation of four participants to each sequence, while A/B variants were balanced within sequences, producing 16 study playlists.}\edit{This produced 16 study playlists that balanced topic order and video-version exposure while preserving a consistent self-guided viewing format.} The videos used in the study can be found in Appendix~\ref{app:videos}.

For the latter two videos in each participant's assigned sequence, participants also completed brief pre- and post-video quizzes targeting concepts covered in the instructional material. \edit{All quizzes were completed on the study computer. We limited the quizzes to two of the four videos to keep the approximately 90-minute study session manageable and reduce participant burden. We selected the latter two videos, rather than the first two, to introduce a delay between the pre-video quizzes administered at the beginning of the session and the corresponding instructional videos. This delay reduced the likelihood that participants would remember specific pre-quiz questions and intentionally search for their answers while watching the videos.} These assessments served as a secondary learning-related measure. In addition to reinforcing the task as a genuine learning activity rather than passive viewing, they allowed us to examine whether participants who reported higher engagement during video viewing also tended to show greater learning gains. Because prior work has linked engagement to improved learning outcomes, such associations provide convergent evidence that the probe captures an educationally meaningful aspect of momentary engagement, while not constituting a direct or exhaustive validation of the construct~\cite{chi2014icap,pintrich1990motivational,kulsoom2022review}. The quizzes were designed to focus on video-specific content while minimizing the influence of general background knowledge. The quiz instruments are included with the released Figshare dataset alongside the multimodal sensor data and other study materials.\footnote{The Figshare release includes a plaintext file listing all quiz questions and answer choices, with correct answer choices marked.}

\subsection{Engagement Measurement}
\label{sec:selfreport}

Because engagement during learning is only partially observable from external behavior, we used repeated in-situ self-report as the primary reference signal. This choice is consistent with prior work that treats brief momentary self-reports as a practical way to capture temporally localized subjective states during ongoing activity while reducing reliance on retrospective recall~\cite{shiffman2008ecological,de2024measuring}. In our study, we used self-report not as an exhaustive measure of engagement, but as a low-burden operationalization of one attentional aspect of momentary engagement. \edit{Throughout this paper, we refer to the recorded responses as \emph{engagement-probe ratings}. These ratings capture participants' perceived difficulty sustaining attention during the preceding minute and should not be interpreted as direct ground-truth measurements of engagement in its entirety.}

Following the framing introduced in Section~\ref{sec:temporal engagement}, we operationalized this aspect of engagement through participants' perceived difficulty maintaining attention. \edit{During scale training, participants were presented with the exact item: ``How difficult was it to pay attention during the last minute of the lecture?'' They responded using a five-point ordinal scale. Only the endpoints were verbally anchored: a rating of 1 indicated that attention was completely automatic and effortless, whereas a rating of 5 indicated a heavy, conscious struggle to keep up with the lesson. Ratings 2--4 represented intermediate values and were not assigned additional verbal anchors. Participants could instead press X to report an external distraction or interruption; these responses were excluded from the prediction labels. The complete instructions are shown in Appendix~\ref{app:prompt}.} Higher ratings therefore indicated greater perceived difficulty sustaining attention. \edit{The engagement-probe ratings directly represent perceived attention difficulty. Although this experience is related to attentional engagement, we do not assume that it is equivalent to engagement or that the two form a simple inverse relationship. For example, a learner may remain engaged with demanding content while simultaneously reporting that sustaining attention requires substantial effort. Conversely, attention may feel easy even when the learner is not deeply cognitively or emotionally engaged.} \edit{The item was intended to capture the subjective effort required to remain focused, rather than the amount or intensity of concentration. This distinction motivated our use of ``difficulty paying attention'' instead of a concentration-focused item: concentration may be intense during both productive engagement and effortful struggle, whereas our item asks specifically about the experienced difficulty of maintaining focus.} This formulation was chosen to capture an attentional dimension of engagement that could be reported quickly and repeatedly during self-guided video learning without requiring lengthy reflection or interrupting the task with a more burdensome assessment~\cite{de2024measuring}.

The decision to use a single-item probe was driven by the need to balance temporal resolution with participant burden. Because our goal was to study fine-grained fluctuations in engagement over time, labels needed to be collected frequently enough to align with short segments of instructional video. At the same time, repeated multi-item questionnaires would have increased disruption, added fatigue, and risked altering the learning experience itself. This concern is consistent with prior work on temporally repeated assessments in cognitive tasks, such as n-back paradigms, where probes must remain brief so that the measurement itself does not substantially interfere with the ongoing task~\cite{owen2005n}. A brief single-item probe therefore provided a practical compromise, allowing us to obtain dense temporal labels while minimizing interference with ongoing video viewing~\cite{dejonckheere2022assessing,song2023examining}. \edit{Prior work on single-item ecological momentary assessment supports the methodological feasibility of collecting brief, repeated subjective reports, but it does not establish that our item measures the complete engagement construct. We therefore interpret the probe as a pragmatic measure of perceived attention difficulty and recognize its narrower construct coverage as a limitation.}

To improve consistency in how participants used the response scale, each participant completed a short practice phase before the main task. During this phase, participants viewed an example clip and responded to sample probes so that they could become familiar with the reporting format and the intended interpretation of the scale. \edit{During the main task, the full question and scale anchors were not shown again. Instead, the interface displayed only the abbreviated cue ``Report engagement now,'' and participants responded using the scale they had learned and practiced.} \edit{We selected an approximately one-minute probing interval as a tradeoff between obtaining temporally localized ratings and limiting participant burden. Prior video-viewing research has sampled momentary cognitive engagement using probes approximately one minute apart~\cite{levordashka2023measuring}, while educational-video research has administered thought probes every 60 seconds to measure mind wandering~\cite{murphy2023effect}. In our study, probes were placed at natural stopping points in the instructional content rather than at rigidly fixed timestamps. Consequently, the exact intervals varied around one minute, but the interruptions occurred at less abrupt points in the videos.}

\input{tables/devices}

\subsection{Sensing Setup}
\label{sec:sensing}

We designed the sensing setup to capture multiple physiological and behavioral signals that may reflect engagement-related processes such as autonomic arousal, attentional demand, cognitive effort, and physical restlessness~\cite{boucsein2012electrodermal,berntson1997heart,conrad2021measuring,hutt2017gaze}. 
At the same time, we intentionally included devices with different sensing capabilities and levels of deployment burden so that later analyses could examine tradeoffs between richer multimodal sensing and more practical configurations~\cite{bustos2022wearables,mach2022assessing,schroeder2023scoping}. Participants wore or used all devices concurrently during the learning task, and we time-synchronized the devices before each session to support temporally aligned analysis across modalities.

To capture motion-related signals, we recorded inertial measurements from devices worn on the head and ear. These streams were included to provide information about movement patterns that may co-vary with attentional state or physical restlessness during video viewing. Specifically, we used the inertial measurement units (IMUs) embedded in the Muse S Athena headband (Interaxon Inc.) and the eSense earable~\cite{kawsar2018esense}, sampled at 52~Hz and 50~Hz, respectively.

\input{figure/study_setup}

To capture cardiovascular activity, we collected signals from multiple devices that differ in form factor and signal type. Heart rate and heart-rate-variability-related measures have been widely linked to arousal, workload, and attentional demand, making them a plausible source of information for momentary engagement estimation~\cite{berntson1997heart,mach2022assessing}. We therefore recorded ECG using a Polar H10 chest strap (Polar Electro Oy, Kempele, Finland) sampled at 130~Hz, and photoplethysmography (PPG) using the $\tau$-Ring~\cite{tang2025ring} and the Muse S Athena, sampled at 25~Hz and 64~Hz, respectively. In addition, the Microsoft Band 2 (Microsoft Corporation, Redmond, WA, USA) provided heart-rate measurements at 2~Hz. These devices provided complementary cardiac measurements while also enabling later comparison between configurations that differ in fidelity and wearability.

We also collected electrodermal activity (EDA) using the Microsoft Band 2 (Microsoft Corporation, Redmond, WA, USA), sampled at 5~Hz. Because fluctuations in skin conductance reflect sympathetic nervous system activity, EDA provides an additional signal related to arousal and attentional or affective responses during learning~\cite{boucsein2012electrodermal}. In addition, we recorded EEG using the Muse S Athena (Interaxon Inc.), sampled at 256~Hz, to capture neural activity that may reflect aspects of cognitive state during instruction~\cite{conrad2021measuring,pope1995biocybernetic}. EEG was \edit{included in} the main modality comparison in our analyses despite its higher deployment burden relative to lighter-weight sensing options.

In addition to the physiological and motion streams, we collected webcam-based eye-tracking data using Beam Eye Tracker software at 30~Hz through Logitech Brio 101 Full HD 1080p camera. Eye tracking was also \edit{included in} the main modality comparison, as gaze behavior provides a complementary view of visual attention during instructional video viewing~\cite{hutt2017gaze,hutt2019automated}. Although webcam-based eye tracking differs in form factor from the wearable signals, we included it because it represents a practically relevant sensing approach for self-guided computer-based learning settings.

A summary of all devices, signal types, placements, and sampling characteristics is provided in \autoref{tab:sensors}. \autoref{fig:study-setup} provides an overview of the study setup, including the wearable device placement on the participant, the computer-based learning environment, and the screen views during normal video watching and engagement reporting periods.

\subsection{Study Procedure}
\label{sec:procedure}

\begin{figure}[t]
    \centering
    \includegraphics[width=0.9\linewidth]{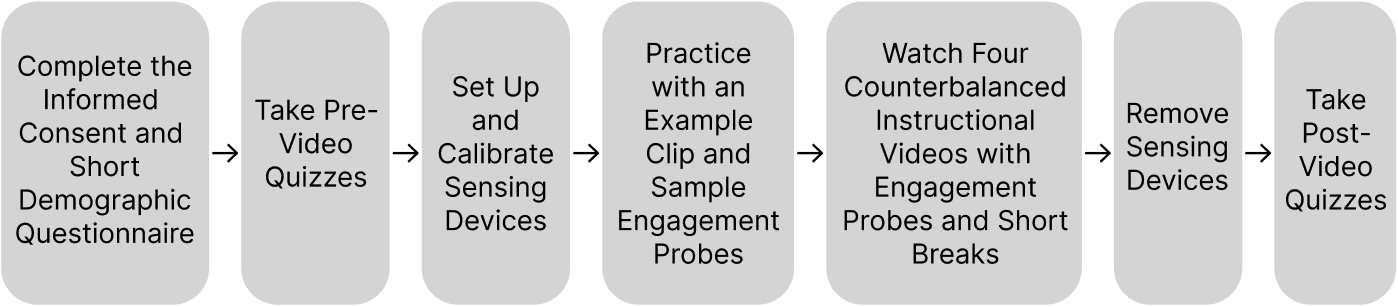}
    \caption{\edit{Overview of the user study procedure.}}
    \label{fig:user_study_process}
\end{figure}

\edit{Each study session lasted approximately 90 minutes and Figure~\ref{fig:user_study_process} summarizes the complete study procedure.} Each session began with informed consent, a short demographic questionnaire, and the \edit{computer-based} pre-video quizzes for the subset of assigned instructional videos used in the learning-gain analysis. After these initial materials were completed, the sensing devices were fitted and calibrated. Participants then completed a brief practice phase in which they watched an example clip and responded to sample engagement probes. This step was intended to familiarize them with the reporting format and the interpretation of the response scale before the main data collection began. \edit{Participants were not given any special restrictions on head, hand, or body movement and were allowed to move naturally while completing the computer-based learning task.}

Participants then completed the main learning task by watching four instructional videos presented in a counterbalanced order.
During video viewing, engagement probes were presented at approximately one-minute intervals. To reduce disruption to the learning experience, however, probes were not inserted at rigidly fixed timestamps. Instead, they were scheduled at natural stopping points in the instructional content whenever possible, so that the interruption would feel less abrupt and responses could more naturally reflect the immediately preceding segment of the lecture. \edit{When a probe appeared, the video paused and the screen displayed the abbreviated cue ``Report engagement now.'' Participants then reported their experience during the preceding minute using the scale learned during training, after which playback resumed.}

Short breaks were offered between videos to reduce fatigue over the course of the session. \edit{To minimize interruptions to the video-viewing procedure and preserve a continuous learning and sensing protocol, the post-video quizzes were not administered immediately after each corresponding video. Instead, after all four videos had been completed, the sensing devices were removed and participants completed the computer-based post-video quizzes for the latter two videos used in the learning-gain analysis.}
\edit{Across this procedure, the study generated temporally aligned self-reports, multimodal sensor streams, probe-level metadata, and quiz responses for constructing labeled windows and interpreting the engagement probe.} We provide additional technical details about the released \projectname{} dataset, including synchronization metadata and study materials, in Appendix~\ref{app:data-collection-system}.

\begin{table*}[t]
\centering
\small
\caption{\edit{Comparison of representative engagement and attention datasets. Physio. denotes physiological or wearable sensing. Restricted indicates that access requires registration or approval.}}
\label{tab:dataset_comparison}
\resizebox{\textwidth}{!}{\begin{tabular}{lccccccc}
\toprule
\edit{\textbf{Dataset}} & \edit{\textbf{Learning Setting}} & \edit{\textbf{Scale}} & \edit{\textbf{Label Source}} & \edit{\textbf{Label Resolution}} & \edit{\textbf{Physio.}} & \edit{\textbf{EEG/Eye}} & \edit{\textbf{Public Access}} \\
\midrule
\edit{DAiSEE~\cite{gupta2016daisee}} & \edit{Self-guided e-learning} & \edit{112 participants; 9,068 clips} & \edit{External raters} & \edit{10 s} & \edit{--} & \edit{--} & \edit{\checkmark} \\
\edit{EngageWild~\cite{kaur2018prediction}} & \edit{Educational videos} & \edit{78 participants; 16.5 h} & \edit{External raters} & \edit{Video level} & \edit{--} & \edit{--} & \edit{Restricted} \\
\edit{CMOSE~\cite{wu2024cmose}} & \edit{Online classes} & \edit{12,193 segments} & \edit{External raters} & \edit{$\sim$10 s} & \edit{--} & \edit{--} & \edit{--} \\
\edit{n-Gage~\cite{gao2020n}} & \edit{High-school classrooms} & \edit{23 participants; 4 weeks} & \edit{In-situ self-report} & \edit{Class level} & \edit{\checkmark} & \edit{--} & \edit{\checkmark} \\
\edit{Reading the Room~\cite{disalvo2022reading}} & \edit{University lectures} & \edit{11 participants; 44 observations} & \edit{Observers} & \edit{Continuous spans} & \edit{\checkmark} & \edit{--} & \edit{--} \\
\edit{DIPSER~\cite{marquez2025dipser}} & \edit{University classrooms} & \edit{54 participants; 51.3 h} & \edit{Expert + retrospective} & \edit{Change points} & \edit{\checkmark} & \edit{--} & \edit{\checkmark} \\
\edit{BBBD~\cite{madsen2026brain}} & \edit{Lab educational videos} & \edit{178 participants; $\sim$110 h} & \edit{Condition + post-video} & \edit{Video level} & \edit{\checkmark} & \edit{\checkmark} & \edit{\checkmark} \\
\edit{\textbf{EduGage}} & \edit{\textbf{Self-guided instructional videos}} & \edit{\textbf{16 participants; $\sim$12 h}} & \edit{\textbf{In-situ self-report}} & \edit{\textbf{$\sim$1 min}} & \edit{\checkmark} & \edit{\checkmark} & \edit{\checkmark} \\
\bottomrule
\end{tabular}}
\end{table*}

\subsection{\projectname{} Dataset}
\edit{We release the \projectname{} dataset as a resource for studying momentary engagement in self-guided video learning. The dataset contains 16 participants, 64 video-viewing sessions, 699 probe-aligned windows, and approximately 12 hours of synchronized multimodal recordings. Each window links the sensor data preceding an engagement probe to the corresponding self-report label, video identifier, segment timing, contextual metadata, and quiz responses where available. We also provide synchronization and preprocessing metadata, participant-based evaluation splits, and study materials to support reproducible model comparison, modality-ablation analysis, and studies of lightweight sensing configurations. The dataset is available at \url{https://doi.org/10.6084/m9.figshare.32145994}.}

\subsubsection{\edit{Dataset Organization and Documentation}}
\edit{The public release is organized to support both dataset inspection and benchmark reproduction. The Figshare archive provides participant-session data using cleaned public participant identifiers, while the accompanying repository documents the expected raw-data layout, sensor-file naming conventions, timestamp conventions, preprocessing stages, window construction, and fixed participant-grouped splits. This organization gives researchers two reuse entry points: they can inspect or reprocess the raw multimodal streams when studying synchronization, feature extraction, or alternative labels, or they can use the provided preprocessing and benchmark code to evaluate new models and reduced-sensor configurations under the same participant-independent protocol used in this paper.}

\subsubsection{\edit{Dataset and Code Availability}}
\edit{The raw dataset is hosted on Figshare, and the accompanying preprocessing, modeling, and benchmark code is available in the public GitHub repository at \url{https://github.com/ZikangLeng/EduGage}. The repository supports raw-data discovery, timestamp repair, probe-aligned window construction, label extraction, participant-grouped split generation, and the benchmark evaluations reported in this paper. The released study materials include the stimulus-video metadata, engagement-probe instructions, demographic questionnaire, and quiz instruments.}

\subsubsection{\edit{Dataset Quality and Validation Checks}}
\edit{We assess dataset quality and validity through several complementary checks. Figure~\ref{fig:label-distribution} reports the distribution of engagement-probe ratings across the 699 supervised windows, showing the expected imbalance toward lower attention difficulty. Appendix~\ref{app:engagement-measure-validation} shows that videos with higher average attention difficulty were associated with lower normalized quiz learning gains, providing evidence that the ratings align with quiz-based learning outcomes. Appendix~\ref{app:data-collection-system} documents where each stream's timestamps came from, how streams were aligned, and the known synchronization limitations, while Appendix~\ref{app:participant-grouped-cv} describes the fixed participant-grouped splits used to avoid participant-level leakage and preserve label coverage across folds. Finally, the benchmark evaluation in Section~\ref{sec:experiments} tests whether the released sensing streams contain predictive information beyond label priors and shared video-time response patterns.}

\subsubsection{\edit{Dataset Rationale and Comparison}}
\edit{Table~\ref{tab:dataset_comparison} situates the \projectname{} dataset relative to representative engagement and attention datasets. Vision-centered datasets such as DAiSEE~\cite{gupta2016daisee}, EngageWild~\cite{kaur2018prediction}, and CMOSE~\cite{wu2024cmose} contain substantially larger collections, but primarily rely on external raters to infer engagement from observable behavior. Classroom datasets such as n-Gage~\cite{gao2020n}, Reading the Room~\cite{disalvo2022reading}, and DIPSER~\cite{marquez2025dipser} provide more naturalistic learning settings and longer observation periods, but use class-level self-reports, observer annotations, or retrospective labels. BBBD~\cite{madsen2026brain} provides a larger collection of neural, physiological, and ocular signals during educational-video viewing, but its reference measures are primarily based on experimentally defined attention conditions and post-video assessments.}

\edit{In contrast, \projectname{} combines self-guided instructional-video learning with repeated in-situ self-reports that are temporally aligned with short multimodal sensing windows. Its primary contribution is therefore not its scale, but the approximately minute-level alignment between synchronized physiological, neural, motion, and eye-tracking signals and participants' perceived difficulty sustaining attention. This design supports the study of within-session fluctuations that cannot be directly captured by externally observed behavior or video- and class-level ratings. At the same time, the reported label represents a specific attentional aspect of engagement rather than the complete multidimensional engagement construct.}

%% file: tables/participants.tex
\begin{table}[t]
\centering
\caption{Participant characteristics for the analyzed sample ($N=16$).
}
\label{tab:participants}
\begin{tabular}{l l}
\toprule
\textbf{Characteristic} & \textbf{Count} \\
\midrule
Gender: Male & 9 \\
Gender: Female & 7 \\
\midrule
Area of study: Computing & 13 \\
Area of study: Other engineering/sciences & 3 \\
\midrule
Weekly online learning: 0 hours & 2 \\
Weekly online learning: 1--3 hours & 7 \\
Weekly online learning: 4--7 hours & 4 \\
Weekly online learning: 8--14 hours & 3 \\
\bottomrule
\end{tabular}
\end{table}

%% file: tables/devices.tex
\begin{table}[t]
\centering
\caption{Wearable sensing devices and signals collected during the study. 
}
\label{tab:sensors}
\begin{tabular}{l l l l}
\toprule
\textbf{Device} & \textbf{Modality} & \textbf{Location} & \textbf{Sampling Rate} \\
\midrule
$\tau$-Ring & PPG / Temperature / IMU & Finger & 25 Hz \\
Microsoft Band 2 & EDA / HR & Wrist & 5 Hz (EDA), 2 Hz (HR) \\
eSense & IMU & Ear & 50 Hz \\
Polar H10 & ECG & Chest & 130 Hz \\
Muse S Athena & EEG / PPG / IMU & Head & 256 Hz (EEG), 64 Hz (PPG), 52 Hz (IMU) \\
Beam Eye Tracker & Eye tracking & Webcam & 30 Hz \\
\bottomrule
\end{tabular}
\end{table}

%% file: figure/study_setup.tex
\begin{figure*}[t]
    \centering

    \begin{subfigure}[t]{0.53\textwidth}
        \centering
        \includegraphics[width=\linewidth]{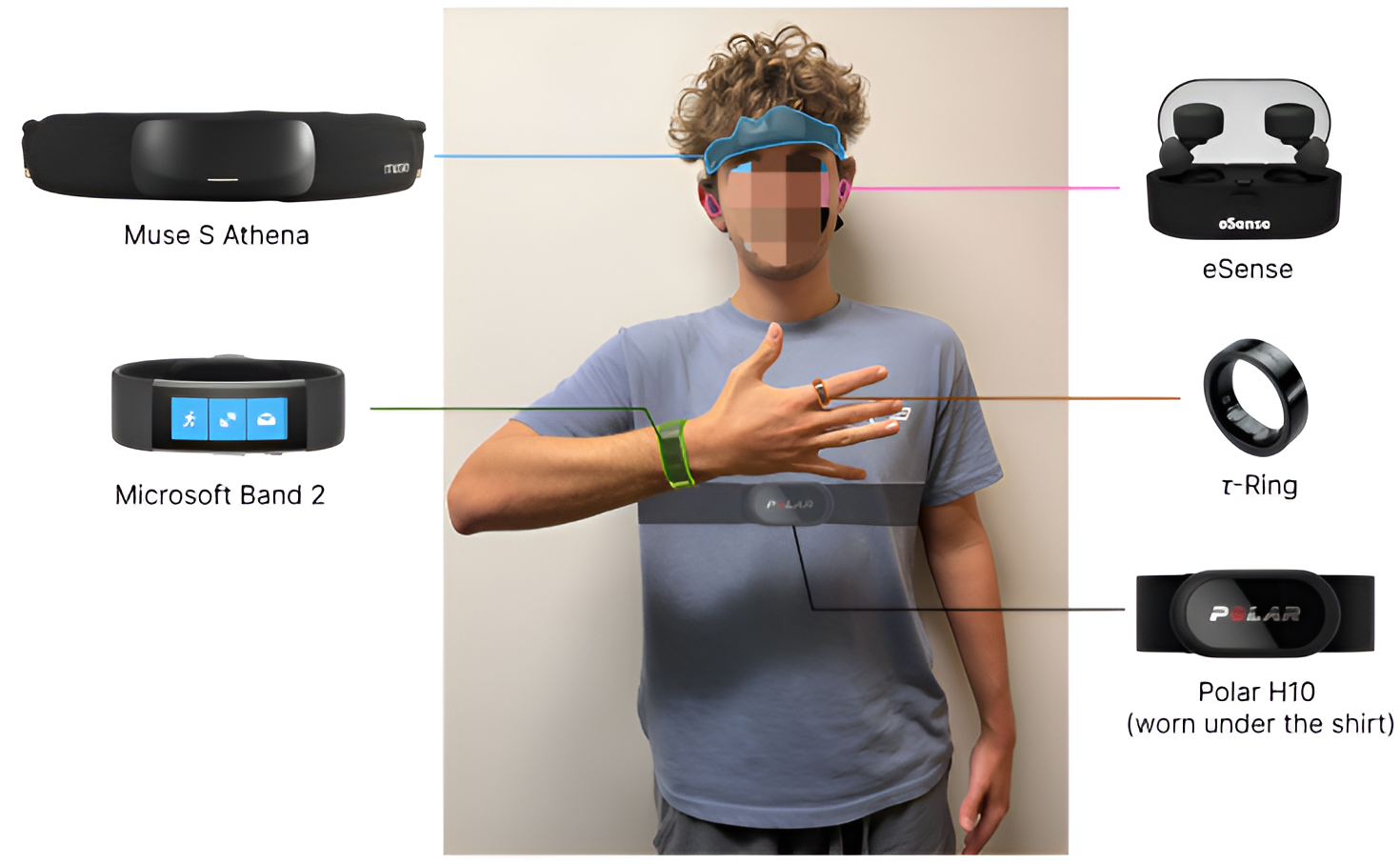}
        \caption{Wearable sensor placement.}
        \label{fig:wearable-placement}
    \end{subfigure}
    \hfill
    \begin{subfigure}[t]{0.44\textwidth}
        \centering
        \includegraphics[width=\linewidth]{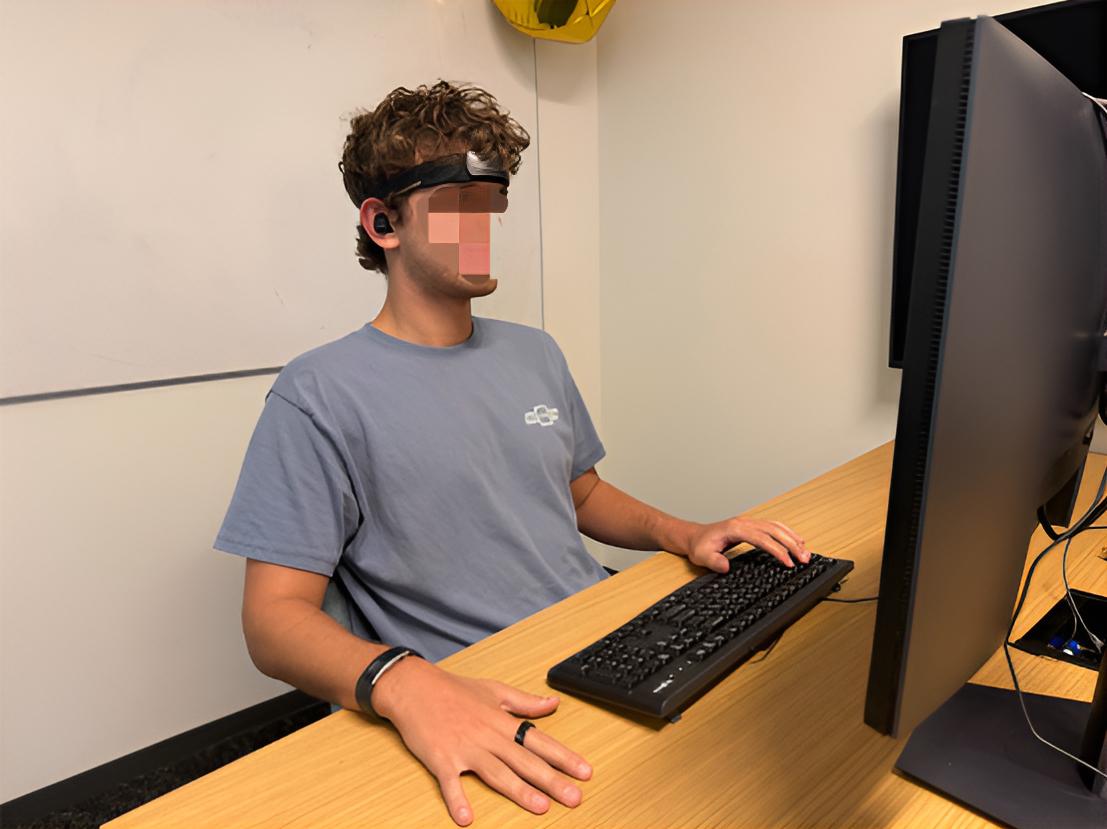}
        \caption{Computer-based video-learning setup.}
        \label{fig:learning-setup}
    \end{subfigure}

    \vspace{0.8em}

    \begin{subfigure}[t]{\textwidth}
        \centering
        \includegraphics[width=0.9\linewidth]{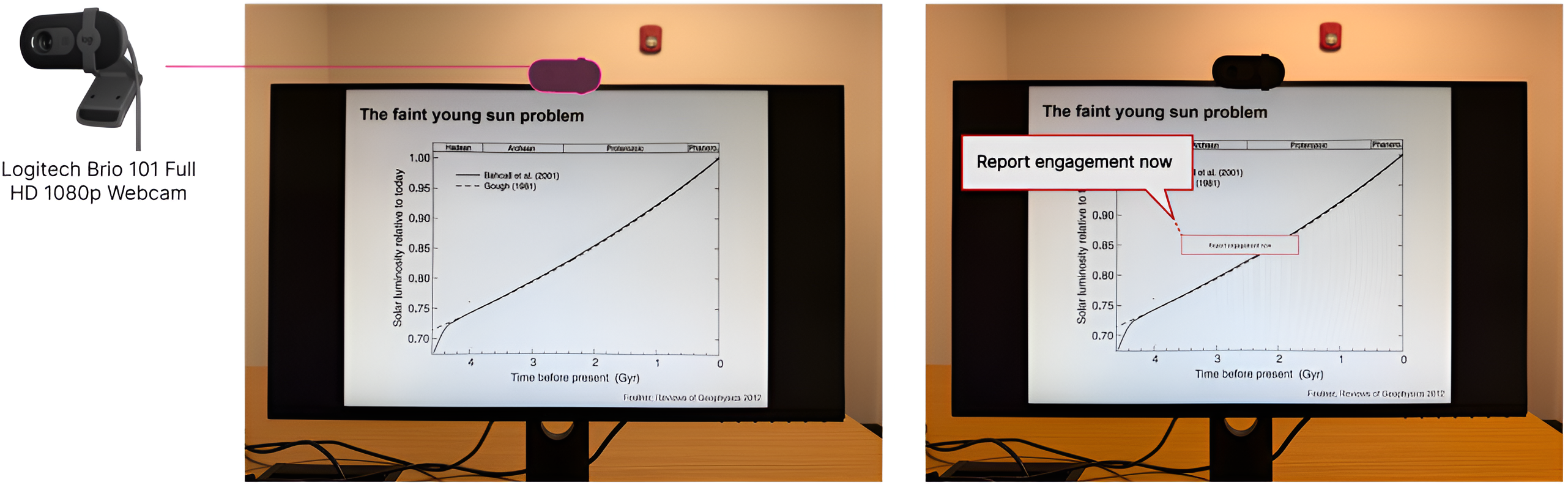}
        \caption{Eye-tracking and in-situ engagement reporting interface. Left: normal video-watching period; right: engagement probe period.}
        \label{fig:eye-tracking-reporting}
    \end{subfigure}

    \caption{Overview of the study setup. Participants watched instructional videos at a desktop workstation while wearing multiple sensors on the head, ears, wrist, finger, and chest. A webcam mounted above the monitor supported eye-tracking. \edit{During each probe, the video paused and the screen displayed the abbreviated cue ``Report engagement now.'' Participants responded using the 1--5 engagement-probe scale introduced during training shown in Appendix~\ref{app:prompt}.} This procedure enabled synchronized collection of sensor signals and self-report ratings.}
    \label{fig:study-setup}
\end{figure*}

%% file: sections/experiment/main.tex
\section{\edit{Benchmark Evaluation and Dataset Validation}}
\label{sec:experiments}

\edit{We evaluate the \projectname{} dataset through a participant-independent benchmark and three complementary analyses. First, we compare sensor-free, statistical, sequence-based, foundation-model, LLM-based, and modality-aware approaches under the same participant-grouped four-fold cross-validation protocol. We then screen all combinations of sensing streams and compare the best-performing combination at each subset size, compare practical cardiac-sensing configurations under matched conditions, and evaluate contemporaneous estimation and forecasting across approximately one- to three-minute horizons. Together, these analyses characterize overall predictive performance, sensing burden, practical cardiac-sensing tradeoffs, and temporal predictability. Detailed model descriptions, implementation settings, and complete benchmark results are provided in Appendix~\ref{app:benchmark-evaluation-details}.}

\subsection{Compact Benchmark Summary}
\label{sec:benchmark_summary}

\begin{table*}[t]
\centering
\small
\caption{\edit{Compact summary of the principal benchmark results. Results are reported as the mean $\pm$ standard deviation across the four participant-grouped folds. Lower MAE is better, whereas higher accuracy and Macro-F1 are better. The complete benchmark is reported in Appendix~\ref{app:benchmark-evaluation-details}.}}
\label{tab:compact_benchmark}
\setlength{\tabcolsep}{4pt}
\renewcommand{\arraystretch}{1.08}
\resizebox{\textwidth}{!}{\begin{tabular}{lcccc}
\toprule
Model
& MAE $\downarrow$
& Within-1 Acc. (\%) $\uparrow$
& \edit{Binary Acc. (\%) $\uparrow$}
& Binary Macro-F1 (\%) $\uparrow$ \\
\midrule

Peer Window Mean ($\pm 5$ s)
& \edit{$1.05 \pm 0.12$}
& \edit{$53.47 \pm 4.24$}
& \edit{$63.49 \pm 8.01$}
& \edit{$50.84 \pm 5.31$} \\

Random Forest
& \edit{$1.02 \pm 0.16$}
& \edit{$76.37 \pm 9.38$}
& \edit{$58.71 \pm 13.44$}
& \edit{$55.68 \pm 12.79$} \\

DeepConvLSTM with Self-Attention
& \edit{$0.96 \pm 0.29$}
& \edit{$78.07 \pm 10.95$}
& \edit{$66.81 \pm 12.26$}
& \edit{$58.37 \pm 14.89$} \\

Concat Embeddings + Linear Head
& \edit{$1.50 \pm 0.12$}
& \edit{$54.19 \pm 2.90$}
& \edit{$53.83 \pm 4.82$}
& \edit{$51.26 \pm 3.99$} \\

LLM-Based Few-shot
& \edit{$1.39 \pm 0.35$}
& \edit{$60.14 \pm 15.92$}
& \edit{$49.72 \pm 12.25$}
& \edit{$43.47 \pm 12.47$} \\

Modality-Aware Model
& \cellcolor{green!20}\edit{$0.80 \pm 0.20$}
& \cellcolor{green!20}\edit{$83.18 \pm 6.78$}
& \cellcolor{green!20}\edit{$70.96 \pm 10.17$}
& \cellcolor{green!20}\edit{$62.90 \pm 8.84$} \\

\bottomrule
\end{tabular}}
\end{table*}

\edit{Table~\ref{tab:compact_benchmark} reports one representative from each benchmark family. Complete results for all evaluated models are provided in Appendix~\ref{app:benchmark-evaluation-details}.} \edit{The modality-aware model achieved the lowest MAE of $0.80$, indicating the most accurate prediction of learners' self-reported difficulty sustaining attention. It also achieved a Within-1 accuracy of $83.18\%$ and a binary Macro-F1 of $62.90\%$, compared with an MAE of $1.05$, a Within-1 accuracy of $53.47\%$, and a binary Macro-F1 of $50.84\%$ for the task-informed Peer Window Mean baseline. These results suggest that the synchronized sensing streams contain predictive information beyond shared video- and time-dependent response patterns. However, the remaining prediction error and moderate classification performance also demonstrate the difficulty of estimating this subjective attentional experience from physiological and behavioral signals.}

\subsection{Modality and Device Configuration Analysis}
\label{sec:modality_analysis}

\edit{To understand the contribution of different sensing streams under deployment constraints, we conduct a modality configuration tradeoff study, since practical wearable sensing systems may not have access to every modality and some sensors impose greater burden, cost, or noise.}

\subsubsection{Experimental Settings}

\edit{We explored all $2^{11}-1=2{,}047$ nonempty combinations of the 11 sensing streams. For each candidate configuration, we train and evaluate the same model using only the selected modalities while holding context features constant, allowing performance changes to reflect the simplified sensing setup.}

\subsubsection{Results}

\begin{table*}[t]
\centering
\caption{\edit{Performance of best sensor configurations containing one to 11 streams. Values are mean $\pm$ standard deviation across four participant-based folds; accuracy and F1 are percentages.}}
\label{tab:subset_optuna_v1_classmae}
\small
\setlength{\tabcolsep}{2pt}
\begin{tabular*}{\textwidth}{@{}p{0.31\textwidth}@{\extracolsep{\fill}}ccccc@{}}
\toprule
\edit{Sensor subset} &
\edit{Streams} &
\edit{Class MAE $\downarrow$} &
\edit{Within-1 (\%) $\uparrow$} &
\edit{Binary Acc (\%) $\uparrow$} &
\edit{Macro-F1 (\%) $\uparrow$} \\
\midrule

\edit{All modalities present} &
\edit{11} &
\colorbox[HTML]{caebc0}{\edit{$0.798 \pm 0.198$}} &
\edit{$83.18 \pm 6.78$} &
\edit{$70.96 \pm 10.17$} &
\edit{$62.90 \pm 8.84$} \\

\edit{All except ECG} &
\edit{10} &
\edit{$0.846 \pm 0.134$} &
\edit{$81.38 \pm 7.83$} &
\edit{$71.19 \pm 10.30$} &
\edit{$62.55 \pm 6.45$} \\

\edit{All except ECG, eIMU} &
\edit{9} &
\edit{$0.828 \pm 0.180$} &
\edit{$81.49 \pm 9.06$} &
\edit{$70.20 \pm 9.25$} &
\edit{$62.67 \pm 7.98$} \\

\edit{All except MPPG, eIMU, RT} &
\edit{8} &
\edit{$0.810 \pm 0.163$} &
\edit{$82.87 \pm 6.71$} &
\edit{$72.07 \pm 9.07$} &
\edit{$66.75 \pm 6.71$} \\

\edit{All except MPPG, MIMU, eIMU, RPPG} &
\edit{7} &
\edit{$0.814 \pm 0.142$} &
\edit{$83.43 \pm 5.77$} &
\edit{$72.79 \pm 6.84$} &
\edit{$69.20 \pm 4.20$} \\

\edit{EEG, EDA, MIMU, HR, RPPG, RIMU} &
\edit{6} &
\edit{$0.801 \pm 0.168$} &
\edit{$83.13 \pm 9.61$} &
\edit{$72.57 \pm 9.11$} &
\edit{$64.50 \pm 12.50$} \\

\edit{EEG, EDA, eIMU, HR, RT} &
\edit{5} &
\edit{$0.804 \pm 0.153$} &
\colorbox[HTML]{caebc0}{\edit{$85.86 \pm 4.72$}} &
\colorbox[HTML]{caebc0}{\edit{$74.44 \pm 8.97$}} &
\edit{$69.91 \pm 7.03$} \\

\edit{EEG, ECG, MPPG, EDA} &
\edit{4} &
\edit{$0.830 \pm 0.207$} &
\edit{$81.04 \pm 11.84$} &
\edit{$70.25 \pm 14.41$} &
\edit{$65.23 \pm 9.64$} \\

\edit{EEG, MPPG, HR} &
\edit{3} &
\edit{$0.890 \pm 0.139$} &
\edit{$81.25 \pm 9.70$} &
\edit{$66.76 \pm 11.13$} &
\edit{$58.66 \pm 9.43$} \\

\edit{ECG, EDA} &
\edit{2} &
\edit{$0.923 \pm 0.168$} &
\edit{$79.32 \pm 11.68$} &
\edit{$66.36 \pm 11.77$} &
\edit{$60.51 \pm 13.68$} \\

\edit{ECG} &
\edit{1} &
\edit{$0.847 \pm 0.092$} &
\edit{$85.63 \pm 5.49$} &
\edit{$74.26 \pm 7.41$} &
\colorbox[HTML]{caebc0}{\edit{$70.54 \pm 6.64$}} \\

\bottomrule
\end{tabular*}

\vspace{2pt}
\parbox{\textwidth}{\footnotesize\edit{MPPG = Muse PPG;
MIMU = Muse IMU; eIMU = eSense IMU; RPPG = Ring PPG;
RIMU = Ring IMU; RT = Ring Temperature.}}
\end{table*}

\edit{Table~\ref{tab:subset_optuna_v1_classmae} compares configurations containing one to 11 sensor streams. The full 11-stream configuration achieves the lowest Class MAE ($0.798 \pm 0.198$), closely followed by the six-stream configuration containing EEG, EDA, Muse IMU, HR, Ring PPG, and Ring IMU ($0.801 \pm 0.168$). The five-stream configuration containing EEG, EDA, eSense IMU, HR, and Ring Temperature achieves the highest Within-1 accuracy ($85.86 \pm 4.72\%$) and binary accuracy ($74.44 \pm 8.97\%$), while ECG alone achieves the highest binary Macro-F1 ($70.54 \pm 6.64\%$).}

\edit{Performance does not change monotonically with the number of streams: smaller configurations sometimes achieve similar or better observed results than larger ones. Because the configurations at adjacent subset sizes were selected independently, differences between rows should not be interpreted as the effect of adding or removing a single stream. The preferred configuration therefore depends on the performance metric and the practical cost of its sensors.}

\subsection{\edit{Cardiac Sensing Across Body Locations}}
\label{sec:cardiac_location_analysis}

\edit{Cardiac activity can be measured using several wearable sensing configurations, but their predictive value may differ because sensor placement affects signal quality, susceptibility to motion artifacts, wearer comfort, and deployment burden. Understanding these differences is important for determining whether engagement estimation requires a relatively intrusive chest-worn ECG sensor or can rely on lower-burden devices worn on the wrist, head, or finger. We therefore independently train models using chest ECG, wrist HR, head PPG, and finger PPG, and additionally evaluate a cardiac-fusion configuration containing all four streams.
}

\begin{table*}[t]
\centering
\caption{\edit{Attention-difficulty prediction using cardiac sensing configurations at different body locations. Results are mean $\pm$ standard deviation across the four participant-based folds.}}
\label{tab:cardiac_location_prediction}
\small
\setlength{\tabcolsep}{5pt}
\begin{tabular}{@{}llcccc@{}}
\toprule
\edit{Input} & \edit{Location} & \edit{Class MAE $\downarrow$} &
\edit{Within-1 Acc. (\%) $\uparrow$} &
\edit{Binary Acc. (\%) $\uparrow$} &
\edit{Binary Macro-F1 (\%) $\uparrow$} \\
\midrule
\edit{ECG} & \edit{Chest} &
\colorbox[HTML]{caebc0}{\edit{$0.847 \pm 0.092$}} &
\colorbox[HTML]{caebc0}{\edit{$85.63 \pm 5.49$}} &
\colorbox[HTML]{caebc0}{\edit{$74.26 \pm 7.41$}} &
\colorbox[HTML]{caebc0}{\edit{$70.54 \pm 6.64$}} \\
\edit{HR} & \edit{Wrist} &
\edit{$1.045 \pm 0.135$} &
\edit{$73.52 \pm 5.09$} &
\edit{$55.12 \pm 10.03$} &
\edit{$49.98 \pm 11.62$} \\
\edit{PPG} & \edit{Head} &
\edit{$1.010 \pm 0.157$} &
\edit{$76.30 \pm 11.46$} &
\edit{$58.84 \pm 15.06$} &
\edit{$44.26 \pm 5.92$} \\
\edit{PPG} & \edit{Finger} &
\edit{$1.043 \pm 0.111$} &
\edit{$73.17 \pm 7.62$} &
\edit{$61.22 \pm 5.16$} &
\edit{$50.09 \pm 7.53$} \\
\edit{Cardiac fusion} & \edit{Multiple} &
\edit{$0.928 \pm 0.233$} &
\edit{$78.94 \pm 10.89$} &
\edit{$64.96 \pm 14.46$} &
\edit{$51.49 \pm 9.49$} \\
\bottomrule
\end{tabular}
\end{table*}

\edit{Table~\ref{tab:cardiac_location_prediction} shows that chest ECG provides the strongest predictive performance across all four metrics. Combining all cardiac streams does not improve upon chest ECG alone, suggesting that the additional cardiac measurements do not provide sufficient complementary information under this setting.}

\edit{Ring PPG appears in several of the selected multisensor combinations in Table~\ref{tab:subset_optuna_v1_classmae}, whereas chest ECG performs best when the cardiac sensing configurations are evaluated individually in Table~\ref{tab:cardiac_location_prediction}. The two analyses address different questions: performance of sensor combinations and performance of individual cardiac sensing configurations.}

\subsection{\edit{Temporal Predictability Across Reporting Horizons}}
\label{sec:temporal_horizons}

\edit{To examine how the frequency of self-report collection relates to the temporal predictability of engagement, we evaluated contemporaneous estimation and direct forecasting at progressively longer horizons. For each source time $t$, the models received the same 44-second sensor window immediately preceding $t$ and predicted the engagement rating at $t$, $t+1$, $t+2$, or $t+3$, corresponding to approximately zero, one, two, and three minutes into the future. Separate models were trained for each horizon. 
Future sensor data and intermediate engagement ratings were not provided to the forecasting models.} \edit{To ensure that differences were not caused by changing sample availability, all horizons were evaluated on the same data for which complete multimodal data and valid targets were available at every horizon. We additionally evaluated a carry-forward baseline that used the engagement rating observed at $t$ as the prediction for each future target. This baseline measures the extent to which forecasting performance can be explained by temporal persistence in participants' self-reports.}

\begin{table*}[t]
\centering
\caption{\edit{Contemporaneous estimation and direct forecasting from the same 44-second source window, together with carry-forward baselines for future horizons. Results are reported as mean $\pm$ standard deviation across the four participant-grouped folds.}}
\label{tab:temporal_horizons}
\small
\setlength{\tabcolsep}{2pt}
\begin{tabular}{@{}llcccc@{}}
\toprule
\edit{Target} &
\edit{Method} &
\edit{Class MAE $\downarrow$} &
\edit{Within-1 Acc. (\%) $\uparrow$} &
\edit{Binary Acc. (\%) $\uparrow$} &
\edit{Binary Macro-F1 (\%) $\uparrow$} \\
\midrule

\multicolumn{6}{@{}l}{\edit{\textit{Direct multimodal models}}} \\

\edit{$t+0$} &
\edit{Contemporaneous estimation} &
\edit{\best{0.753 \pm 0.173}} &
\edit{$86.25 \pm 6.62$} &
\edit{$74.50 \pm 10.10$} &
\edit{$68.88 \pm 7.34$} \\

\edit{$t+1$ ($\sim$1 min)} &
\edit{Direct forecasting} &
\edit{$0.756 \pm 0.135$} &
\edit{\best{88.40 \pm 5.31}} &
\edit{\best{76.43 \pm 6.26}} &
\edit{\best{71.31 \pm 4.11}} \\

\edit{$t+2$ ($\sim$2 min)} &
\edit{Direct forecasting} &
\edit{$0.812 \pm 0.152$} &
\edit{$84.27 \pm 10.20$} &
\edit{$73.58 \pm 9.76$} &
\edit{$68.07 \pm 7.93$} \\

\edit{$t+3$ ($\sim$3 min)} &
\edit{Direct forecasting} &
\edit{$0.844 \pm 0.099$} &
\edit{$80.17 \pm 4.65$} &
\edit{$71.32 \pm 8.94$} &
\edit{$67.68 \pm 7.92$} \\

\midrule
\multicolumn{6}{@{}l}{\edit{\textit{Carry-forward baselines}}} \\

\edit{$t+1$ ($\sim$1 min)} &
\edit{Last observed rating} &
\edit{\best{0.615 \pm 0.106}} &
\edit{\best{91.33 \pm 4.55}} &
\edit{\best{78.90 \pm 5.88}} &
\edit{\best{74.40 \pm 3.19}} \\

\edit{$t+2$ ($\sim$2 min)} &
\edit{Last observed rating} &
\edit{$0.740 \pm 0.116$} &
\edit{$86.43 \pm 5.66$} &
\edit{$74.97 \pm 3.47$} &
\edit{$69.38 \pm 2.64$} \\

\edit{$t+3$ ($\sim$3 min)} &
\edit{Last observed rating} &
\edit{$0.696 \pm 0.050$} &
\edit{$86.46 \pm 4.57$} &
\edit{$77.75 \pm 3.29$} &
\edit{$73.14 \pm 1.20$} \\

\bottomrule
\end{tabular}
\end{table*}

\edit{As shown in Table~\ref{tab:temporal_horizons}, the one-minute direct forecast performs similarly to contemporaneous estimation. Class MAE changes only from 0.753 at $t+0$ to 0.756 at $t+1$, while Within-1 accuracy, binary accuracy, and binary Macro-F1 are slightly higher at $t+1$. Performance becomes weaker at the longer horizons: relative to $t+1$, Class MAE increases by 0.056 at $t+2$ and by 0.088 at $t+3$, while Within-1 accuracy decreases by 4.13 and 8.23 percentage points, respectively.}

\edit{The carry-forward baseline outperforms the directly trained sensor model at every future horizon, demonstrating substantial temporal persistence in the self-reported ratings. Consequently, these results do not identify an optimal self-report collection interval. Because all participants experienced the same approximately one-minute prompting protocol, determining the effect of probe frequency would require a study that directly compares different randomized collection intervals.}

%% file: sections/discussion/main.tex
\section{Discussion}

\subsection{Fine-Grained Engagement Estimation Is Feasible but Inherently Noisy}

Our results suggest that wearable and contextual signals can support fine-grained engagement estimation during self-guided video learning. Although exact self-reported scores remain difficult to predict, the models capture meaningful temporal structure in learners' moment-to-moment attention difficulty. This is important because engagement is not static across a learning session; it changes as instructional content unfolds, difficulty varies, and attention fluctuates over time~\cite{conrad2021measuring,kizilcec2013deconstructing}.

\edit{A representative prediction trace is provided in Appendix~\ref{app:prediction-trace}. The example illustrates that the model can capture coarse temporal trends, even when exact point-level agreement with each self-reported rating remains difficult.}

\edit{Prediction discrepancies are expected, given the nature of the task.} Engagement is a partially latent construct that cannot be directly observed from physiology or behavior alone~\cite{henrie2015measuring}. Moreover, our target is not objective engagement itself, but self-reported difficulty in sustaining attention during the preceding video segment. These reports are subjective and may be influenced by response style, scale interpretation, recent memory, and probe timing. Physiological and behavioral signals are also indirect: they may reflect attention, arousal, cognitive effort, fatigue, prior knowledge, video difficulty, movement, or distraction, which may not always align synchronously with self-report.

These findings highlight both the promise and limitation of fine-grained engagement sensing. Wearable sensing can approximate moment-to-moment engagement dynamics, but perfect agreement with self-report should not be expected. In practical learning systems, such models may be most useful for identifying broader temporal patterns, such as sustained increases in attention difficulty or video segments where learners may benefit from reflection, review, or support.

\subsection{Modeling Lessons from Limited Multimodal Data}

The modeling results show that fine-grained engagement estimation remains a data-limited and subtle prediction problem. Although multimodal sensing provides a rich set of physiological and behavioral signals, these signals do not map cleanly onto self-reported engagement~\cite{booth2023engagement,gao2020n}. Under this setting, simpler statistical models remain competitive with more complex deep learning architectures. This suggests that compact feature summaries can provide useful robustness when the number of participants and labeled windows is limited, while deep learning models may require larger datasets, repeated sessions, or stronger personalization to fully benefit from raw temporal dynamics.

The results also suggest that multimodal engagement modeling should account for modality heterogeneity rather than relying only on simple early fusion. Different sensors capture different aspects of the learning process, and their reliability can vary across participants, windows, and recording conditions. A modality-aware fusion design is therefore useful because it allows the model to preserve modality-specific structure before combining evidence across streams. This is especially important in wearable sensing, where missing data, noise, and variable signal quality are common~\cite{van2024mitigating}.

Our foundation-model and LLM-based baselines further highlight that engagement prediction is not a plug-and-play transfer problem. Pretrained biosignal and time-series encoders did not automatically produce stronger representations for this task, likely because their pretraining objectives and source domains are not directly aligned with momentary engagement estimation. Similarly, the LLM-based few-shot baseline could reason over structured sensor summaries, but struggled with fine-grained ordinal distinctions. Together, these findings suggest that future progress may require engagement-specific representation learning, better personalization, and models that explicitly account for the temporal and subjective nature of the labels.

\subsection{Relating Benchmark Performance to Real-World Learning Scenarios}

\edit{Prior work provides numerical context for our benchmark, although differences in prediction targets and evaluation protocols prevent direct performance rankings. n-Gage reported an MAE of $0.602$ for an overall engagement score rescaled to a 1--5 range, compared with our MAE of $0.80$~\cite{gao2020n}. However, n-Gage predicted a class-level measure aggregated across behavioral, emotional, and cognitive engagement, whereas we predict repeatedly reported attention-difficulty ratings for individual video segments.}

\edit{Probe-based mind-wandering studies more closely resemble the temporal structure of our task. \citet{hutt2017gaze} formulated prediction as binary classification between mind wandering and being on task. Using 30-second gaze windows preceding thought probes during lecture viewing, they reported an F1 score of $47\%$ for the mind-wandering class, compared with $30\%$ for chance~\cite{hutt2017gaze}. In authentic high-school classrooms, subsequent work reported mind-wandering-class F1 scores of $59\%$ offline and $40\%$ live, compared with $24\%$ for chance~\cite{hutt2019automated}. Our binary Macro-F1 of $62.90\%$ provides encouraging evidence for attention-difficulty prediction, but it is not directly equivalent to these results because the studies use different labels, class distributions, sensing settings, and validation protocols. These findings nevertheless suggest potential applications such as recommending difficult segments for optional review, while the reduction from offline to live classroom performance highlights the challenges of real-world deployment.}

\subsection{Deployment Tradeoffs for Wearable Engagement Sensing}

For engagement-aware learning systems to be useful outside the laboratory, the sensing setup must be practical for everyday use~\cite{yan2022scalability}. Our study used a broad multimodal configuration to examine which signals may be informative, but a real learner watching videos at home, in a library, or during a short study session is unlikely to wear multiple devices or tolerate extensive setup. The key deployment question is therefore not whether more sensors improve performance, but which smaller configuration provides sufficient information with minimal burden~\cite{booth2023engagement}.

\edit{The configuration analyses show that more sensing streams do not consistently improve every metric. The 11-stream configuration has the lowest Class MAE, the five-stream configuration has the highest Within-1 and binary accuracy, and ECG alone has the highest binary Macro-F1. Chest ECG also performs best in the cardiac-sensing comparison, while combining the cardiac streams does not improve upon ECG alone. These findings suggest selecting sensors according to the intended performance metric and deployment burden. Because the cardiac configurations differ in sensing principle, device, sampling rate, and body location, that comparison does not isolate the effect of anatomical placement alone.}

A realistic configuration for laptop-based video learning would likely combine one low-burden behavioral signal with one lightweight physiological signal. For example, webcam-based gaze or head-pose features could be paired with a wristband or smart ring that captures PPG, heart rate, temperature, or EDA. For mobile, headset-based, or future smart-glasses scenarios, earables or glasses-based sensing may be more natural~\cite{bustos2022wearables}. In contrast, EEG headbands and chest ECG straps are better viewed as research-grade sensors rather than default choices for everyday learning.

Overall, wearable engagement sensing should aim for low-burden, approximate feedback rather than clinical-grade measurement. Even imperfect estimates may be useful for detecting sustained attention difficulty, identifying video segments that commonly challenge learners, or triggering low-risk support such as review prompts or reflection cues~\cite{booth2023engagement}. Future systems should therefore use the least intrusive available signals, handle missing modalities, and treat engagement estimates as uncertain indicators rather than definitive judgments. 

\edit{The relevant stakeholders also differ in their goals and authority. Researchers in wearable sensing, affective computing, and learning analytics may use the dataset to study multimodal modeling and reduced-sensor configurations. Potential system users include learners, instructors, content designers, and educational institutions. Learner-controlled estimates could support personal reflection, optional review suggestions, or other low-risk adaptive prompts, while aggregated patterns across learners could help instructors and content designers identify video segments that consistently warrant clarification. These uses are meaningfully different from institution-controlled monitoring: engagement estimates should not be treated as objective evidence for grading, discipline, or other high-stakes decisions.}

\edit{Gaze and physiological streams are sensitive data, and models may enable inferences about attention or other internal states that learners did not intend to disclose. Responsible use therefore requires explicit and revocable consent, data minimization, secure storage, restricted access, clear retention policies, and communication of model uncertainty. Whenever possible, systems should process the least sensitive signals needed and return feedback to the learner rather than expose continuous individual-level traces to an institution. Learners should also be able to opt out without penalty. Even with these safeguards, the data and resulting models should not be used for covert surveillance or punitive evaluation.}

\subsection{Supporting Benchmarking and Reuse in Multimodal Learning Analytics}
\edit{As a primary contribution of this work}, we release the \projectname{} dataset as a resource for studying momentary engagement in self-guided video learning. While prior engagement research and learning analytics resources have supported important progress, many focus on classroom interaction, platform logs, session-level outcomes, or coarser behavioral annotations~\cite{disalvo2022reading,gao2020n}. In contrast, our dataset provides synchronized multimodal sensor streams together with repeated segment-level self-reports collected during video-based learning. This design makes it possible to study engagement as a temporally situated and fluctuating process rather than as a single aggregate property of an entire learning session.

The dataset can support several lines of future research. First, it provides a benchmark for evaluating alternative models for fine-grained engagement estimation, including feature-based models, temporal sequence models, foundation-model transfer approaches, and multimodal fusion methods. Second, because the dataset contains multiple sensing streams, it enables analysis of modality contribution, sensor selection, and robustness to missing or noisy wearable data. Third, the repeated self-reports and \edit{quiz-based} learning measures allow researchers to examine how momentary attention difficulty relates to broader learning behaviors and outcomes. Finally, the inclusion of multiple videos and participants enables initial studies of personalization and cross-content generalization.

\edit{For example, a researcher developing a lower-burden engagement-estimation system for laptop-based learning could use only the released webcam-based eye-tracking and wrist-worn EDA and heart-rate streams, train and evaluate the model using the provided participant-grouped splits, and compare this reduced sensing configuration against the full multimodal benchmark. This would allow the researcher to study the tradeoff between predictive performance and sensing burden without reproducing the original multimodal data collection.}

By releasing this dataset, we aim to support more reproducible and comparable research on multimodal engagement sensing. Recent work in learning analytics has emphasized that open datasets are important for reproducibility, collaboration, and trust, yet raw datasets are still rarely released alongside learning analytics papers~\cite{vsvabensky2026open}. Our dataset helps address this gap for self-guided video learning by providing temporally aligned multimodal data and dense engagement-related labels. At the same time, the dataset should be interpreted in light of its scope: it was collected in a controlled laboratory setting with a specific learning task and participant sample. This controlled design provides synchronized and well-structured data for model development, while future work should extend this line of research to larger, longer-term, and more naturalistic learning environments.

\subsection{Limitations and Future Work}

This study opens several directions for future work. First, our controlled laboratory design allowed us to collect synchronized multimodal sensor streams and dense segment-level self-reports, providing a reliable basis for studying fine-grained engagement estimation. \edit{Our sample of 16 participants supported balanced allocation across the four presentation sequences, but it remains modest for evaluating population-level generalization. Other intensive wearable-engagement studies have involved similarly sized cohorts, including n-Gage with 23 students and Reading the Room with 11 students~\cite{gao2020n,disalvo2022reading}. However, comparable cohort sizes in prior work do not establish statistical sufficiency. We therefore characterize \projectname{} as an intensive repeated-measures dataset comprising 16 participants, 64 video-viewing sessions, and 699 probe-aligned windows. Our participant-grouped evaluation measures generalization to held-out learners within this cohort rather than to the broader student population.} A natural next step is to extend this approach to larger and more diverse learning contexts, including longer-term studies in homes, libraries, classrooms, and other everyday environments where learners use different devices and encounter more natural distractions~\cite{yan2022scalability}.

Second, our engagement probe focused on the perceived difficulty of sustaining attention during the preceding video segment. This low-burden measure enabled frequent temporal sampling without heavily interrupting the learning task~\cite{booth2023engagement}. \edit{However, the frequent in-situ probes may have affected the attentional states they were intended to measure. Pausing the video and asking participants to reflect on their attention could temporarily refocus them on the learning task and alter their attention during the following segment. One alternative is video-stimulated recall, in which participants are shown short video segments after completing the video and asked to retrospectively report their attention during each segment~\cite{pereira2024attention}. Although this approach avoids interrupting the initial viewing experience, it introduces potential memory and retrospective-report biases. Comparing in-situ and retrospective measures across different probing frequencies would help clarify these tradeoffs.} Future work can build on this design by combining attention-difficulty probes with complementary measures, such as interaction logs, pause and replay behavior, post-video reflections, comprehension outcomes, or qualitative interviews. Such multimethod validation would help clarify how momentary attention difficulty relates to broader cognitive, emotional, and behavioral dimensions of engagement~\cite{fredricks2004school}.

Beyond model evaluation, fine-grained engagement sensing can also support new forms of adaptive learning. In future systems, momentary estimates of attention difficulty could be used to provide timely support when learners appear to struggle, such as suggesting a short pause, prompting reflection, recommending that a difficult segment be reviewed, or adapting the pace of instruction. Aggregated over time, these estimates could also provide learners with post-hoc feedback about when they were most challenged or disengaged during a study session. At the instructor or content-design level, engagement traces across learners could help identify video segments that consistently produce difficulty, confusion, or loss of attention, offering evidence for redesigning explanations, adding examples, or restructuring instructional materials.

%% file: sections/conclusion/main.tex
\section{Conclusion}

We presented \edit{the \projectname{} multimodal dataset and benchmark for} fine-grained engagement estimation in self-guided video learning. In this study, 16 college students watched instructional videos while wearing or using multiple physiological and behavioral sensing devices, including PPG, heart rate, EDA, ECG, EEG, IMU, temperature, and webcam-based eye tracking. Participants provided repeated in-situ self-reports of attention difficulty, resulting in the \projectname{} dataset of synchronized multimodal sensor streams, probe-aligned engagement labels, study materials, and learning-related measures.

Using this \projectname{} dataset, we evaluated whether wearable sensing can support momentary engagement estimation and how different modeling and sensing choices affect performance. \edit{As one method within this benchmark, we evaluated a modality-aware model} that learns modality-specific temporal representations and combines them through context-informed gated fusion. Under participant-based cross-validation, \edit{the model} achieved an MAE of 0.80, 83.18\% within-1 accuracy, 70.96\% binary accuracy, and 62.90\% binary Macro-F1, outperforming sensor-free baselines, statistical regressors, deep temporal architectures, foundation-model transfer baselines, and LLM-based few-shot reasoning.

More broadly, our findings suggest that fine-grained engagement estimation is feasible but inherently noisy. Wearable and contextual signals can approximate learners' moment-to-moment engagement. \edit{Our modality analysis shows that smaller sensor configurations can perform well on different metrics, motivating sensor selection according to the intended use and deployment burden.} By releasing the \projectname{} dataset and study materials, we aim to support future work on reproducible, deployable, and fine-grained engagement modeling for self-guided learning.

%% file: sections/ackowledgement/main.tex
\section*{Acknowledgments}
This work was partially supported by the NSF Research Fellowship under Grant No. DGE-2039655.

Any opinion, findings, and conclusions or recommendations expressed in this material are those of the authors(s) and do not necessarily reflect the views of the National Science Foundation.

%% file: sections/appendix/main.tex
\appendix

\input{sections/method/main}

\section{\edit{Benchmark Evaluation Details}}
\label{app:benchmark-evaluation-details}
\input{sections/appendix/benchmark_evaluation_details}

\section{Implementation Details for Modeling Experiments}
\label{app:modeling-details}

\subsection{Input Representation, Labels, and Evaluation Protocol}
\label{app:model-input-representation}
\input{sections/appendix/model_input_representation}

\input{sections/appendix/label_distribution}

\subsection{Supervised Models Trained from Scratch}
\label{app:supervised-model-details}

\subsubsection{Classical Machine Learning Features and Models}
\label{app:ml-feature-details}
\input{sections/appendix/ml_features_models}

\input{sections/appendix/supervised_model_details}

\subsection{Foundation-Model Transfer}
\label{app:foundation-model-transfer}
\input{sections/appendix/foundation_model_transfer_details}

\subsection{\edit{LLM-Based Few-Shot Reasoning}}
\label{app:llm-baseline}
\input{sections/appendix/llm_baseline_details}

\section{Data Collection System}
\label{app:data-collection-system}
\input{sections/appendix/data_collection_system}

\section{Study Materials and Instruments}

\subsection{Engagement Measure Validation}
\label{app:engagement-measure-validation}
\input{sections/appendix/engagement_measure_validation}

\subsection{Demographic Survey}
\label{app:demographics}
\input{sections/appendix/demographics}

\subsection{Stimulus Videos}
\label{app:videos}
\input{sections/appendix/videos}

\subsection{Experience Sampling Prompt}
\label{app:prompt}
\input{sections/appendix/prompt}

\subsection{Comprehension Quizzes}
\label{app:quizzes}
\input{sections/appendix/quizzes}

\section{\edit{Participant-grouped Cross-validation Protocol}}
\label{app:participant-grouped-cv}
\input{sections/appendix/cross_validation_explanation}

\section{\edit{Deployment Footprint and Practical Cost}}
\label{app:deployment-footprint}
\input{sections/appendix/deployment_footprint}

%% file: sections/method/main.tex
\section{Multimodal Engagement Prediction \edit{Model}}
\label{app:multimodal-engagement-model}

\edit{This appendix section presents the multimodal model used in our benchmark} to estimate momentary learner engagement from synchronized sensor streams. 
We use the term \emph{window} to refer to a temporal segment ended on each engagement probe, where a probe is the moment at which the learner provides a self-reported engagement rating. Each probe-aligned window is treated as a single prediction sample and contains modality-specific sensor signals together with video-based contextual metadata. 
The model learns modality-specific latent representations, adaptively weights their contributions using a context-informed gating mechanism, and combines them to produce a final scalar engagement estimate.

\subsection{\edit{Model Overview}}
We model learner engagement at the level of probe-aligned windows extracted from continuous multimodal recordings during instructional video viewing. For each window, sensor streams are aligned to a common timeline, and a modality-specific temporal input is extracted from each available sensing stream over the corresponding time interval. A compact metadata vector describing the window’s relative position within the video is retained as contextual input.

\begin{figure}[t]
    \centering
    \includegraphics[width=\linewidth]{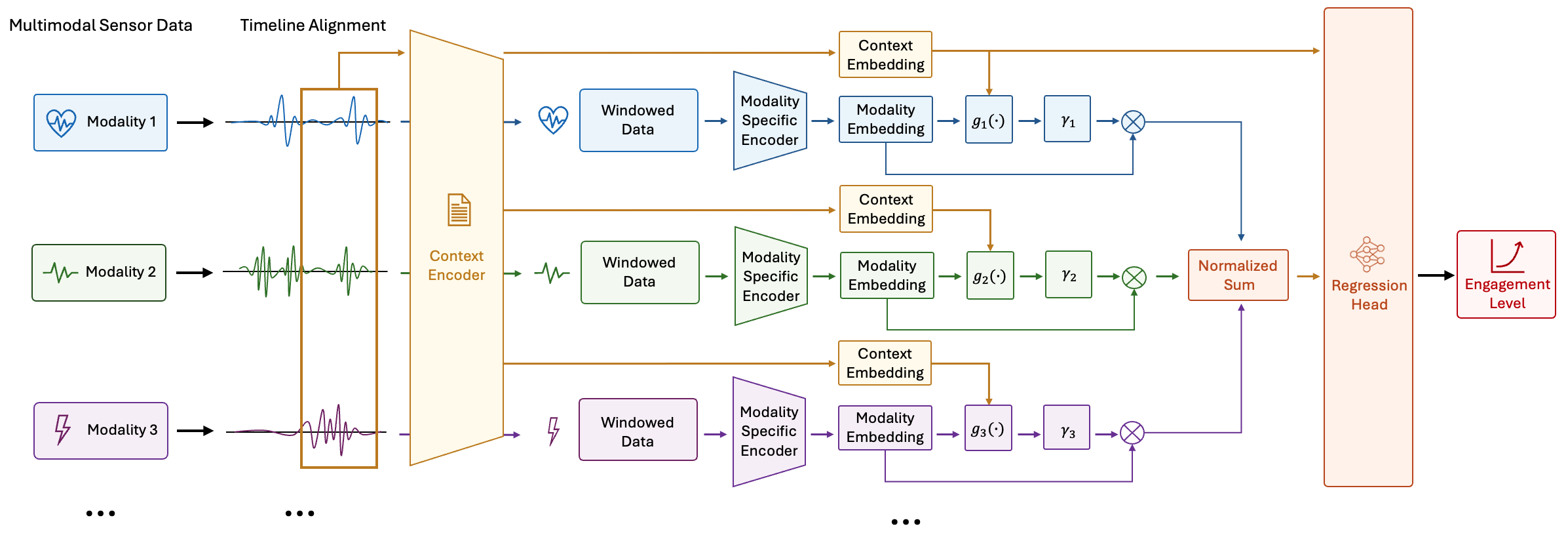}
    \caption{\edit{Overview of the modality-aware model used in the \projectname{} benchmark.} Continuous multimodal sensor streams are aligned on a shared timeline and partitioned into probe-aligned windows with common metadata. For each window, modality-specific temporal inputs are extracted and encoded into latent representations. A context-informed gating module estimates a contribution weight for each modality, and the weighted modality embeddings are fused into a shared representation. The fused representation, together with encoded context information, is then mapped to the final engagement estimate by a regression head.}
    \label{fig:outline}
\end{figure}

Figure~\ref{fig:outline} illustrates the overall architecture. Each modality-specific temporal input is processed independently by a temporal encoder to obtain a latent embedding. A gating module then estimates the importance of each modality conditioned on its embedding and the shared context vector. The modality embeddings are fused through a normalized gated weighted average, and the fused representation is concatenated with the encoded context representation before being passed to a regression head that outputs the predicted engagement level.

This architecture separates three complementary functions: 
\textit{i)} modality-specific temporal representation learning;
\textit{ii)} adaptive modality weighting; and 
\textit{iii)} final predictive mapping. Such a decomposition is especially useful in multimodal engagement modeling, where sensing streams differ in sampling rate, signal quality, and predictive reliability across windows and participants.

\subsection{Input Representation}

\edit{For each probe-aligned learning window and each sensing modality, the model receives a raw temporal input containing the modality's available sensor channels over the window. Because the model operates on native-frequency sensor streams, the number of temporal samples may differ across modalities. Each window is also associated with a shared context vector that summarizes coarse temporal information about the window within the instructional video. Specifically, we use relative progress through the video and a sinusoidal encoding of this progress as the context.}

\subsection{Modality-Specific Encoding}

\edit{Each modality is processed by its own temporal encoder to produce a latent modality embedding.}

\edit{In parallel, the context vector is transformed by a small context encoder, producing the encoded context representation used during gating and final prediction.}

\subsection{Context-Informed Gated Fusion}

Because wearable sensing streams differ in signal quality, sampling rate, missingness, and relevance to engagement-related processes, we use a context-informed gated fusion mechanism that learns a soft contribution weight for each modality in each probe-aligned window. This design follows the broader principle of gated expert weighting and gated multimodal fusion, where learned gates control how strongly different information sources contribute to the final representation \citep{jacobs1991adaptive, arevalo2017gated}.

\edit{For each modality, the model computes a modality-specific gate from the modality embedding and the encoded context vector. Unlike shared gating units that estimate fusion weights from a joint multimodal representation, we use separate gating networks for different modalities. This parameterization allows each sensing stream to learn its own reliability and contribution pattern, which is useful for heterogeneous signals such as physiological, motion, neural, and eye-tracking data. The scalar gate value is obtained using a sigmoid activation and is set to zero when the modality is unavailable for the current window. The final multimodal representation is then computed as a normalized gated average over the available modality embeddings, with a small numerical-stability constant in the denominator.}

This mechanism allows the model to adaptively emphasize modalities that appear more informative for a given window while reducing the influence of noisier or less reliable streams. Because the gates are continuous and differentiable, the model can learn graded modality contributions rather than making hard modality-selection decisions. We interpret the learned gates as model-estimated contribution weights rather than causal measures of modality importance.

\subsection{Engagement Prediction}

\edit{The fused modality representation is concatenated with the encoded context vector and passed to the final prediction head, which outputs the predicted engagement score for the window.}

%% file: sections/appendix/benchmark_evaluation_details.tex
\subsection{Experimental Setup}
\label{sec:experimental_setup}

\subsubsection{Prediction Task}

We formulate momentary engagement estimation as a supervised prediction task over probe-aligned learning windows. Each sample corresponds to one \edit{44-second} multimodal time window extracted immediately before an in-situ engagement probe. \edit{The prediction target is the participant's 1--5 engagement-probe rating in response to the item: \emph{``How difficult was it to pay attention during the last minute of the lecture?''}}

\edit{Higher engagement-probe ratings indicate greater reported difficulty sustaining attention. We treat the responses as an ordered scale and evaluate predictions according to their closeness to the reported value. Although the measure operationalizes one attentional aspect of engagement, we do not interpret it as a complete measure or simple inverse of engagement.}

\subsubsection{Data Processing, Input Representation, and Evaluation Protocol}

Continuous sensor streams are temporally aligned and converted into probe-aligned learning windows of 44 seconds. For each engagement probe, we extract multimodal sensor data from the immediately preceding time interval and retain modality-specific inputs at their native sampling resolutions. 
We additionally compute lightweight contextual metadata describing the relative temporal position of each window within the instructional video. Samples associated with explicit external distraction or interruption are excluded from the prediction task.

To support different modeling approaches, we construct both raw temporal sequence inputs and statistical feature representations from the same aligned windows. Raw temporal sequences are used by the deep learning models, whereas statistical feature vectors are used by classical machine learning baselines and LLM-based approaches. Additional implementation details for feature construction are provided in Appendix \ref{app:ml-feature-details}.

We evaluate generalization using participant-based 4-fold cross-validation, where held-out folds consist of disjoint groups of participants. This grouped evaluation protocol is used instead of leave-one-subject-out validation because not all individual participants exhibit the full range of label values. \edit{We provide additional details and a visual illustration of this evaluation protocol in Appendix~\ref{app:participant-grouped-cv}. }

\edit{Because the target ratings form an ordered 1--5 scale, we evaluate ordinal prediction performance using mean absolute error (MAE) and Within-1 accuracy. MAE measures the average absolute difference between the predicted and reported ratings, in units of the response scale. Within-1 accuracy is the percentage of predictions that differ from the reported rating by no more than one response level. We additionally report binary accuracy and binary Macro-F1 after grouping ratings 1--2 as lower attention difficulty and ratings 3--5 as higher attention difficulty. Binary accuracy measures the percentage of correctly classified windows, whereas Macro-F1 computes the F1 score for each class separately and averages the two scores, giving equal weight to the lower- and higher-attention-difficulty classes. We compute each metric separately for the four held-out participant groups and report the mean and standard deviation across folds.}

\subsection{\edit{Benchmark Settings and Results}}
\label{sec:benchmark_settings_results}

\subsubsection{\edit{Experimental Settings}}
\label{sec:experimental_settings}

\edit{We organize the benchmark settings into sensor-free baselines, task-specific supervised models, foundation-model transfer baselines, and LLM-based reasoning.} For the task-specific supervised setting, we train models directly on the \projectname{} dataset. This setting reflects the standard approach in wearable and biosignal-based engagement recognition, where models learn mappings from labeled sensor windows to engagement estimates. We include two supervised baseline families: feature-based models that operate on hand-crafted statistical summaries, and sequence-based models that learn directly from raw temporal sensor inputs.

\paragraph{Feature-Based ML Baselines}
For the feature-based machine learning baselines, we summarize each sensor channel within the window using statistical features such as distributional and temporal descriptors (see Appendix \ref{app:ml-feature-details}). Features from all available modalities are concatenated into a single early-fused vector, together with encoded contextual metadata. We evaluate Random Forest, Linear Regression, LightGBM, and RBF SVM regressors.

\paragraph{Sequence-Based DL Baselines}
For the sequence-based deep learning baselines, we use raw temporal sensor inputs. Modalities are resampled or padded as needed to form a fixed-length multichannel temporal tensor, and contextual metadata is concatenated with the learned sequence embedding before the final prediction layer. We evaluate DeepConvLSTM~\cite{ordonez2016deep}, DeepConvLSTM with Self Attention~\cite{singh2021DeepConvLSTM}, and TinyHAR~\cite{inproceedings} to test whether end-to-end temporal learning can capture engagement-relevant dynamics beyond statistical summaries. More implementation details are shown in Appendix~\ref{app:supervised-model-details}.

\paragraph{Sensor-Free Baselines}
To quantify whether sensor-based models provide information beyond dataset-level label priors \edit{and task-level response patterns}, we compare against \edit{four} non-sensor baselines: mean, mode, \edit{random sampling from the empirical label distribution, and peer window mean}. \edit{The first three} baselines use only the training-label distribution within each fold and do not access physiological, behavioral, or contextual inputs. The mode and mean baselines produce constant predictions per fold, while the random baseline samples predictions according to the empirical distribution. \edit{The peer window mean baseline incorporates task context by predicting each engagement report from the mean responses of other participants watching the same video within a $\pm 5$-second window. } \edit{The prior-based comparison} provides a lower-bound reference for assessing whether learned models capture meaningful sensor-engagement \edit{relationships, while the peer window mean provides a stronger, task-informed reference for determining whether engagement can be explained by shared responses to the video content rather than participant-specific sensor signals}. Similar prior-based baselines are commonly used in engagement and affective sensing evaluations \cite{di2018unobtrusive, gao2020n}.

\paragraph{Foundation-Model Transfer Baselines}
\edit{We evaluate foundation-model transfer by extracting frozen window-level embeddings from pretrained biosignal and time-series encoders, including PulsePPG, ECGFounder, NeuroLM, NormWear, and MOMENT~\cite{saha2025pulse,li2024electrocardiogram,jiang2024neurolm,luo2024toward,goswami2024moment}. We then train lightweight prediction heads on top of these embeddings, using either direct concatenation or a gated fusion layer. This setting tests whether external pretraining provides useful task-aligned representations for engagement prediction; implementation details are provided in Appendix~\ref{app:foundation-model-transfer}.}

\paragraph{LLM-Based Few-Shot Reasoning}
\edit{We also evaluate a few-shot LLM reasoning baseline inspired by ConSensus~\cite{yoon2026consensus}. This baseline converts engineered sensor features into structured text summaries, retrieves training-fold examples from participants outside the held-out fold, and uses GPT-5-mini with zero-temperature decoding to produce window-level engagement predictions through stream-specific reasoning and lightweight fusion prompts. This setting is intended as a reference comparison for prompt-based multimodal reasoning rather than as an efficient deployment model; details are provided in Appendix~\ref{app:llm-baseline}.}

\paragraph{Modality-Aware Model}
\edit{Finally, we evaluate a modality-aware supervised model that encodes each sensing stream separately, estimates context-informed contribution weights for available modalities, and fuses the weighted modality representations before predicting the engagement score. This design preserves modality-specific structure before cross-modal integration, which is useful when sensors differ in sampling rate, noise level, missingness, and relevance across windows. The full architecture is described in Appendix~\ref{app:multimodal-engagement-model}.}

\subsubsection{\edit{Results}}
\label{sec:modeling_results}

Results are summarized in Table~\ref{tab:supervised_from_scratch_baselines_results}.
\input{tables/supervised_from_scratch_baselines_results.tex}
\edit{The modality-aware model achieved the strongest overall performance among the evaluated models, with an MAE of $0.80$ and a binary Macro-F1 of $62.90\%$. An MAE of $0.80$ means that the predicted rating differed from the participant's reported rating by approximately $0.80$ response levels on average on the 1--5 scale. Its Within-1 accuracy of $83.18\%$ indicates that most predictions were within one response level of the reported rating.}

\edit{The sensor-free baselines establish lower-bound comparisons. The mean baseline performs reasonably on MAE and Within-1 accuracy because the labels are concentrated toward lower attention difficulty, but it performs poorly on Macro-F1 because it does not distinguish the binary states. The peer window mean baseline captures shared video-level response patterns, but remains below the stronger sensor-based models.}

The feature-based statistical regressors provide a competitive but limited baseline. Random Forest, LightGBM, and SVM achieve reasonable MAE and within-1 accuracy, indicating that statistical summaries of wearable signals contain useful engagement-related information. However, their binary Macro-F1 scores remain substantially lower than \edit{the modality-aware model}. This suggests that simple early fusion over hand-crafted features can capture coarse trends in the data, but may struggle to model how the reliability and relevance of different modalities change across windows and participants.

The sequence-based temporal models do not consistently outperform the statistical regressors. DeepConvLSTM performs relatively weakly, while the attention-based variant and TinyHAR improve the results, with \edit{the attention-based variant achieving the strongest binary Macro-F1 among the sequence-based temporal model baselines (58.37\%)}. However, these models still remain below \edit{the modality-aware model}. One likely reason is that directly combining heterogeneous sensor streams into a single temporal representation requires more data to learn stable cross-modal patterns. In our setting, the signals differ substantially in sampling rate, noise level, and physiological meaning, making early temporal fusion difficult under the current data scale.

\edit{The foundation-model transfer baselines also underperform the supervised modality-aware model across all evaluation metrics. Direct concatenation of frozen foundation-model embeddings with video-progress metadata yields limited performance, with an MAE of $1.50$ and a binary Macro-F1 of $51.26\%$. The gated fusion variant does not improve over simple concatenation. This indicates that frozen pretrained embeddings, at least in this setting, do not provide sufficient task-aligned information for fine-grained engagement estimation.}

\edit{The LLM few-shot baseline shows limited performance on the original prediction task, with an MAE of $1.39$. Although the LLM can reason over structured modality summaries, the subtle differences among fine-grained engagement-probe rating levels are difficult to infer from short sensor windows using only in-context examples rather than task-specific training. The gap remains clear under the binary formulation: the LLM baseline achieves a binary accuracy of $49.72\%$ and a binary Macro-F1 of $43.47\%$. Overall, these results indicate that few-shot LLM-based reasoning is not sufficient for reliable fine-grained engagement recognition in its current form.}

\edit{Compared with these baselines, the modality-aware model preserves modality-specific structure before cross-modal integration. Each sensing stream is first encoded separately, and the resulting representations are then combined through context-informed gated fusion. This design better matches the sensing setting, where a modality may be informative in one window but noisy or less relevant in another. The improvement over statistical regressors, sequence-based temporal architectures, foundation-model transfer baselines, and LLM-based few-shot reasoning suggests that fine-grained engagement estimation benefits from task-specific modality-aware integration.}

\subsection{\edit{Representative Prediction Trace}}
\label{app:prediction-trace}

\begin{figure}[t]
    \centering
    \includegraphics[width=0.7\linewidth]{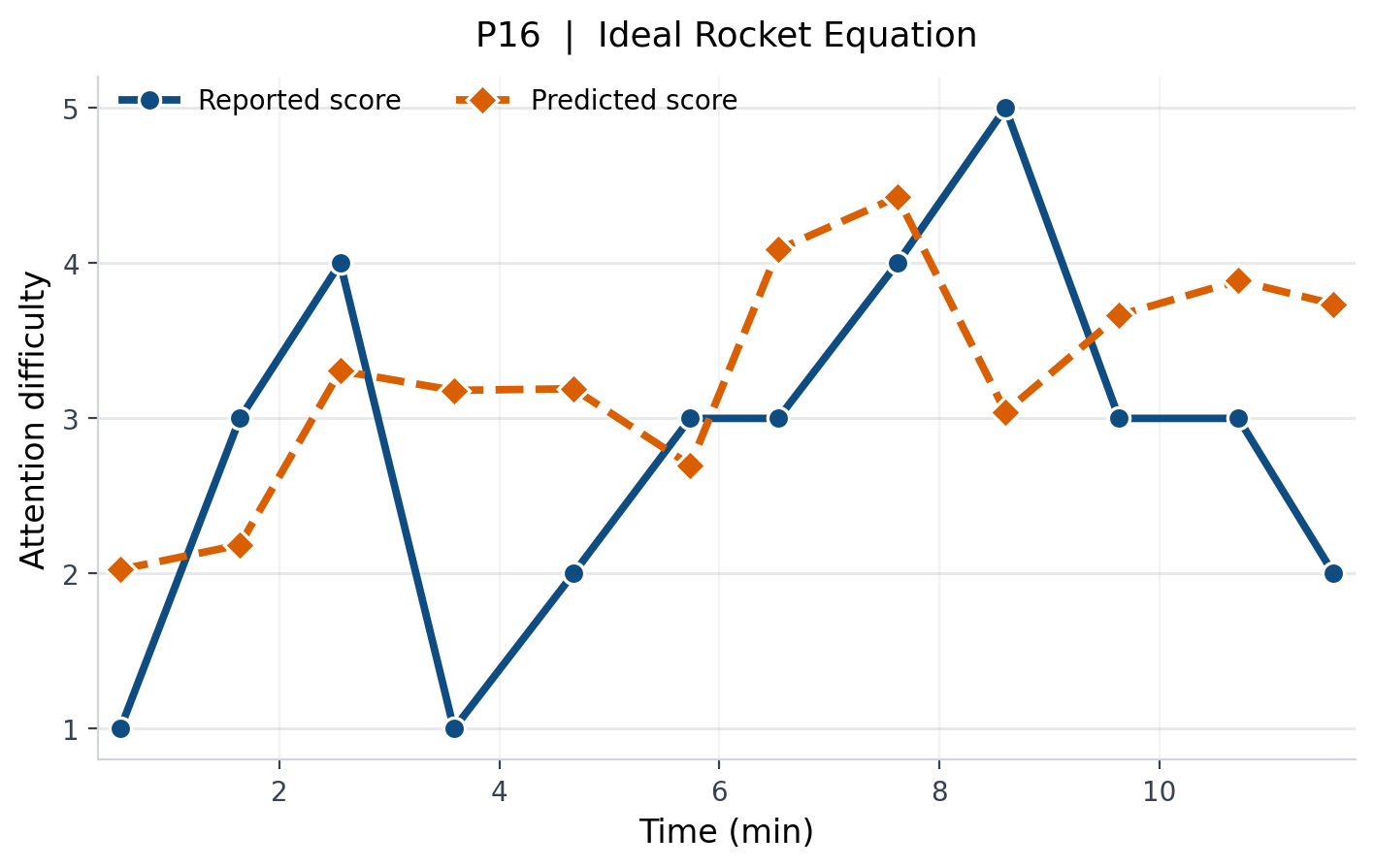}
    \caption{Example temporal trace for participant 16 while watching the \textit{Ideal Rocket Equation} video. The solid blue curve shows the self-reported attention-difficulty score at each probe-aligned window, while the dashed orange curve shows the model's predicted score. 
    }
    \label{fig:prediction_example}
\end{figure}

Figure~\ref{fig:prediction_example} shows a representative example from one participant and video. The predicted trajectory does not perfectly match the self-reported trajectory, but the two series are partially aligned over time. In particular, predicted scores often follow the general direction of the self-reported scores, suggesting that the model captures coarse temporal trends rather than exact point-level labels.

%% file: tables/supervised_from_scratch_baselines_results.tex
\begin{table*}[t]
    \centering

    \caption{\edit{Performance comparison of all benchmark models. Fold-wise means are reported with standard deviations.}
    }
    \label{tab:supervised_from_scratch_baselines_results}

\small
    \resizebox{\textwidth}{!}{\begin{tabular}{@{}lcccc@{}}
        \toprule
        {Model} & MAE $\downarrow$ & Within-1 Acc (\%) $\uparrow$ & Binary Acc (\%) $\uparrow$ & Binary Macro-F1 (\%) $\uparrow$ \\
        \midrule
        \multicolumn{5}{@{}l}{\textit{Sensor-Free Baselines}} \\
        \midrule
        Mean & \edit{0.98 $\pm$ 0.18} & \edit{81.71 $\pm$ 8.72} & \edit{62.11 $\pm$ 13.36} & \edit{37.99 $\pm$ 5.29} \\
        Mode & \edit{1.31 $\pm$ 0.37} & \edit{62.11 $\pm$ 13.36} & \edit{62.11 $\pm$ 13.36} & \edit{37.99 $\pm$ 5.29} \\
        Random Distribution & \edit{1.36 $\pm$ 0.11} & \edit{60.24 $\pm$ 4.79} & \edit{52.80 $\pm$ 4.13} & \edit{50.15 $\pm$ 3.10} \\
        Peer Window Mean ($\pm 5$ s) &
        \edit{1.05 $\pm$ 0.12} &
        \edit{53.47 $\pm$ 4.24} &
        \edit{63.49 $\pm$ 8.01} &
        \edit{50.84 $\pm$ 5.31} \\
        \midrule
        \multicolumn{5}{@{}l}{\textit{ML Statistical Regressors}} \\
        \midrule
        Random Forest & \edit{1.02 $\pm$ 0.16} & \edit{76.37 $\pm$ 9.38} & \edit{58.71 $\pm$ 13.44} & \edit{55.68 $\pm$ 12.79} \\
        Linear Regression & \edit{1.25 $\pm$ 0.21} & \edit{62.86 $\pm$ 8.03} & \edit{55.21 $\pm$ 13.61} & \edit{54.17 $\pm$ 13.65} \\
        LightGBM & \edit{1.04 $\pm$ 0.16} & \edit{75.11 $\pm$ 11.20} & \edit{58.88 $\pm$ 13.02} & \edit{55.17 $\pm$ 12.63} \\
        SVM & \edit{1.07 $\pm$ 0.23} & \edit{75.71 $\pm$ 10.24} & \edit{55.57 $\pm$ 13.89} & \edit{46.12 $\pm$ 16.62} \\
        
        \midrule
        \multicolumn{5}{@{}l}{\textit{Sequence-Based DL Baselines}} \\
        \midrule
        DeepConvLSTM & \edit{1.11 $\pm$ 0.32} & \edit{72.20 $\pm$ 10.34} & \edit{55.37 $\pm$ 12.94} & \edit{50.76 $\pm$ 10.14} \\
        DeepConvLSTM with Self-Attention & \edit{0.96 $\pm$ 0.29} & \edit{78.07 $\pm$ 10.95} & \edit{66.81 $\pm$ 12.26} & \edit{58.37 $\pm$ 14.89} \\
        TinyHAR & \edit{1.16 $\pm$ 0.21} & \edit{72.71 $\pm$ 6.58} & \edit{59.60 $\pm$ 8.18} & \edit{55.13 $\pm$ 9.88} \\
        \midrule
        \multicolumn{5}{@{}l}{\edit{\textit{Foundation-Model Transfer Baselines}}} \\
        \midrule
        \edit{Concat Embeddings + Linear Head} & \edit{1.50 $\pm$ 0.12} & \edit{54.19 $\pm$ 2.90} & \edit{53.83 $\pm$ 4.82} & \edit{51.26 $\pm$ 3.99} \\
        \edit{Gated Fusion} & \edit{1.55 $\pm$ 0.27} & \edit{53.73 $\pm$ 10.35} & \edit{41.58 $\pm$ 15.77} & \edit{39.38 $\pm$ 16.97} \\
        \midrule
        \multicolumn{5}{@{}l}{\edit{\textit{LLM-Based Reasoning Baseline}}} \\
        \midrule
        \edit{LLM-Based Few-shot} & \edit{1.39 $\pm$ 0.35} & \edit{60.14 $\pm$ 15.92} & \edit{49.72 $\pm$ 12.25} & \edit{43.47 $\pm$ 12.47} \\
        \midrule
        Modality-Aware Model & \colorbox[HTML]{caebc0}{\edit{$0.80 \pm 0.20$}} & \colorbox[HTML]{caebc0}{\edit{$83.18 \pm 6.78$}} & \colorbox[HTML]{caebc0}{\edit{$70.96 \pm 10.17$}} & \colorbox[HTML]{caebc0}{\edit{$62.90 \pm 8.84$}} \\
        \bottomrule
        
    \end{tabular}}
\end{table*}

%% file: sections/appendix/model_input_representation.tex
This section describes the common windowing, input channels, and evaluation protocol used by the modeling experiments. Each prediction instance was anchored to an engagement probe. \edit{Although the engagement-probe item asked participants to consider approximately the preceding minute, probes were placed at natural stopping points in the videos rather than at rigidly fixed timestamps. Consequently, the uninterrupted viewing intervals between probes varied, with the shortest interval lasting 44 seconds. We therefore used the final 44 seconds of viewing immediately preceding each probe as the fixed model-input window. This ensured that all prediction samples had the same duration and that no window overlapped with the preceding probe-and-response period. Thus, the approximately one-minute reporting interval and the 44-second modeling window were related but not identical.}

For sequence-based supervised models, each window was represented as
\[
    \mathbf{X}\in\mathbb{R}^{28\times 2200},
\]
corresponding to 28 channels resampled to 50 Hz. Missing streams or missing samples within a channel were represented as zeros after interpolation/resampling. Each fold standardized the raw tensor channel-wise using the training participants only, with means and standard deviations computed across the training windows and time axis.

The 28 input channels were:
\begin{enumerate}[leftmargin=*]
    \item Beam eye/head tracking: screen gaze $x$, screen gaze $y$, and head position $x,y,z$.
    \item Microsoft Band: EDA resistance in kOhms and heart rate in bpm.
    \item Muse PPG: left outer 730 nm optical channel.
    \item Ring: green optical PPG, and three raw temperature channels: two on the inside of the ring, left and right, and one on the outside.
    \item eSense IMU: acceleration $x,y,z$ and gyroscope $x,y,z$.
    \item Polar H10: ECG amplitude.
    \item Muse IMU: acceleration $x,y,z$ and gyroscope $x,y,z$.
    \item Muse EEG: AF7 and AF8. The available Muse EEG electrodes were AF7, AF8, TP9, and TP10, with FPz as reference, following the extended 10--10 EEG positioning system~\cite{SEECK20172070}. Figure~\ref{fig:eeg-channel-placement} shows the retained AF7/AF8 electrodes and FPz reference location. We used AF7 and AF8 and discarded TP9 and TP10 because temporal (mastoid) electrodes are substantially noisier. They are also less robust for participants with long hair, as they require contact with skin above the ear and under the hair, whereas the AF7 and AF8 electrodes are positioned on the forehead.
    \item Metadata: video-progress position within the current video. For example, the fourth window in a 12-window video has metadata value approximately 0.3.
\end{enumerate}

\input{sections/appendix/fig_eeg_channel_placement}

Models were evaluated with participant-based 4-fold cross validation rather than leave-one-subject-out evaluation. This choice was made because not every participant contributed examples for all five engagement labels, and the higher engagement-difficulty labels were comparatively sparse. Grouping participants into four folds allowed each test fold to contain all five labels while preserving participant-level separation between training and testing.

%% file: sections/appendix/fig_eeg_channel_placement.tex
\begin{figure}[t]
    \centering
    \includegraphics[width=0.75\linewidth]{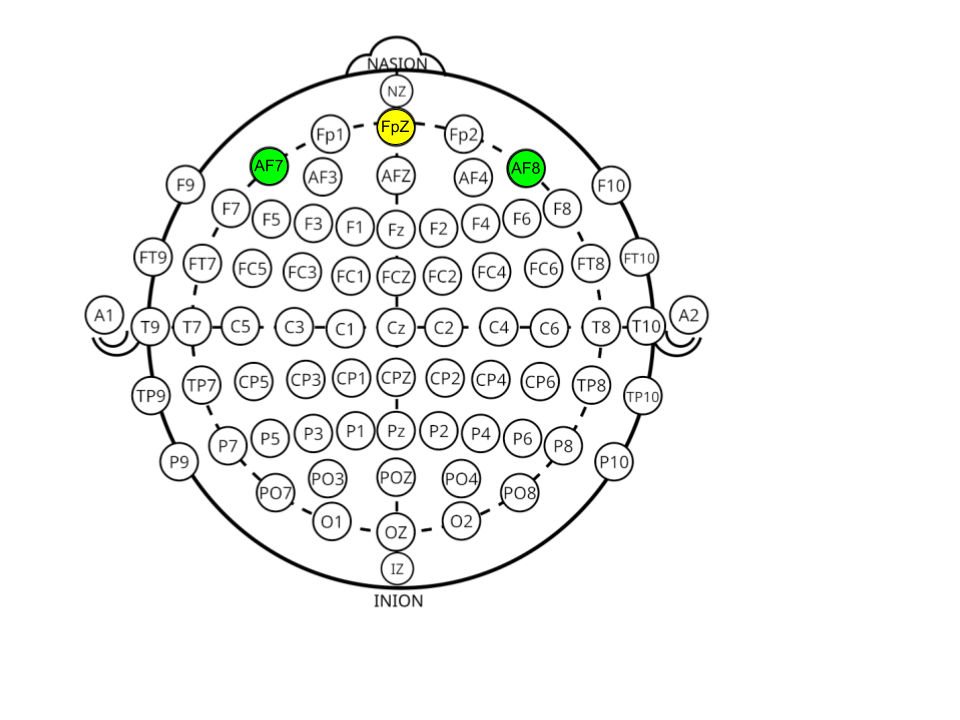}
    \caption{Muse EEG electrode placement in the extended 10--10 system. AF7 and AF8, shaded green, were retained for modeling. FPz, shaded yellow, was used as the reference location.}
    \label{fig:eeg-channel-placement}
\end{figure}

%% file: sections/appendix/label_distribution.tex
Figure~\ref{fig:label-distribution} shows the distribution of the five engagement-probe labels over the \edit{699} supervised windows used in the modeling experiments. \edit{The labels are participants' 1--5 responses to the item, ``How difficult was it to pay attention during the last minute of the lecture?''}, where 1 indicates effortless attention and 5 indicates substantial effort to maintain attention. The distribution is imbalanced toward lower attention difficulty: \edit{labels 1 and 2 account for 434 of 699 windows, or 62.1\% of the dataset. Labels 4 and 5 account for 128 windows, or 18.3\%.}

\input{sections/appendix/fig_label_distribution}

%% file: sections/appendix/fig_label_distribution.tex
\begin{figure}[t]
    \centering
    \input{sections/appendix/label_distribution_barchart}
    \caption{\edit{Distribution of engagement-probe ratings across all 699 supervised windows. Higher ratings indicate greater reported difficulty sustaining attention.}}
    \label{fig:label-distribution}
\end{figure}

%% file: sections/appendix/label_distribution_barchart.tex
\begin{tikzpicture}[x=1.15cm,y=0.018cm]
    \draw[->] (0,0) -- (6.2,0) node[right] {Label};
    \draw[->] (0,0) -- (0,270) node[above] {Windows};
    \foreach \y in {0,50,100,150,200,250} {
        \draw (-0.05,\y) -- (0.05,\y);
        \node[left] at (-0.08,\y) {\scriptsize \y};
        \draw[gray!20] (0,\y) -- (5.8,\y);
    }
    \foreach \x/\h/\pct in {1/233/33.3\%,2/201/28.8\%,3/137/19.6\%,4/70/10.0\%,5/58/8.3\%} {
        \fill[black!65] (\x-0.32,0) rectangle (\x+0.32,\h);
        \node[below] at (\x,-6) {\scriptsize \x};
        \node[above] at (\x,\h+7) {\scriptsize \edit{\h}};
        \node[above] at (\x,\h+22) {\scriptsize \edit{\pct}};
    }
\end{tikzpicture}

%% file: sections/appendix/ml_features_models.tex
The classical machine learning baselines used the same event-trailing windows as the sequence models. Features were computed directly from the preprocessed modality streams at their native sampling rates. Each available channel within a window was summarized independently with eight statistical features.

\paragraph{Statistical Features}

For each channel $c$ with native-rate samples $x_{c,1:T_c}$ in the event-trailing window, we computed the following features:
\begin{enumerate}[leftmargin=*]
    \item \textbf{Mean}: $\frac{1}{T_c}\sum_t x_{c,t}$.
    \item \textbf{Standard deviation}: temporal standard deviation of the channel.
    \item \textbf{Kurtosis}: excess kurtosis, computed as $\frac{\mathbb{E}[(x-\mu)^4]}{\mathbb{E}[(x-\mu)^2]^2}-3$, with zero used when the denominator is numerically zero.
    \item \textbf{Minimum}: $\min_t x_{c,t}$.
    \item \textbf{Maximum}: $\max_t x_{c,t}$.
    \item \textbf{Interquartile range}: $p_{75}-p_{25}$.
    \item \textbf{Difference standard deviation}: standard deviation of the first difference sequence.
    \item \textbf{Slope}: $(x_{c,T_c}-x_{c,1})/(T_c-1)$.
\end{enumerate}

Missing, non-finite, or infinite values were converted to zero or clipped to the range $[-10^6,10^6]$ before model fitting. For the linear and SVM models, features were additionally robust-scaled using training-fold medians and interquartile ranges, then clipped to $[-10,10]$ after scaling.

\paragraph{Classical Models}
All classical baselines were implemented as regressors over ordinal labels encoded as integers 0--4. Predictions were rounded to the nearest integer and clipped to the valid label range before conversion back to labels 1--5. This formulation preserves ordinal distance information and supports mean absolute error as a natural metric.

\begin{table}[t]
\centering
\caption{Classical ML models and implemented hyperparameters.}
\label{tab:ml-model-details}
\small
\begin{tabular}{lp{0.65\linewidth}}
\toprule
Model & Implementation details \\
\midrule
Random forest & \texttt{sklearn.ensemble.RandomForestRegressor}; 500 trees; \texttt{min\_samples\_leaf=2}. \\
Linear regression & Implemented as ridge regression: \texttt{RobustScaler}, post-scaling clip to $[-10,10]$, then \texttt{sklearn.linear\_model.Ridge} with $\alpha=1.0$ and \texttt{solver=lsqr}. \\
LightGBM & \texttt{LGBMRegressor}; squared-error regression objective; 500 trees; learning rate 0.03; 15 leaves; \texttt{min\_child\_samples=10}; subsample 0.9; column subsample 0.8. \\
RBF SVM & \texttt{sklearn.svm.SVR}; \texttt{RobustScaler}, post-scaling clip to $[-10,10]$, RBF kernel, $C=1.0$, and \texttt{gamma=scale}. \\
\bottomrule
\end{tabular}
\end{table}

The classical models used the same participant-based 4-fold cross validation as the other supervised baselines. As above, this was used instead of leave-one-subject-out evaluation because not every participant had all five labels, and the fold design ensured that each test fold contained all five engagement labels. For the binary metric in this implementation, ordinal labels 1--2 were mapped to the lower attention-difficulty class and labels 3--5 were mapped to the higher attention-difficulty class.

%% file: sections/appendix/supervised_model_details.tex
The supervised models trained from scratch consisted of classical machine learning regressors over statistical features and deep sequence models over raw temporal windows. All supervised baselines used participant-based 4-fold cross validation and predicted engagement labels encoded from 0 to 4. Continuous predictions were rounded and clipped to the valid label range before computing five-class MAE, within-one-label accuracy, and binary low/high metrics. Binary labels were rebinned as labels 1--2 for lower attention difficulty and labels 3--5 for higher attention difficulty.

\subsubsection{Deep Sequence Models}
\edit{For each deep sequence model, we ran a small hyperparameter sweep over learning rate, weight decay, and dropout, and report the best-performing configuration by validation MAE.}
\paragraph{DeepConvLSTM}
The DeepConvLSTM baseline follows the convolutional-recurrent architecture introduced for multimodal wearable activity recognition~\cite{ordonez2016deep}. It uses four temporal convolution layers with kernel size $5\times 1$, 64 filters per layer, and ReLU activations, followed by two LSTM layers with 128 hidden units. The final recurrent state is passed through dropout and a scalar regression head trained with unweighted L1/MAE loss.

\paragraph{DeepConvLSTM with Self-Attention}
The attention variant extends the convolutional-recurrent model with a self-attention pooling module for wearable-sensor sequence modeling~\cite{singh2021DeepConvLSTM}. It applies a channel-spanning convolution with three filters, processes the result with a 32-unit LSTM, and pools the recurrent sequence using multi-hop self-attention with attention size 32 and 10 attention hops. The attended representation is passed to a scalar regression head trained with unweighted L1/MAE loss.

\paragraph{TinyHAR}
The TinyHAR baseline follows the lightweight hierarchical attention design for human activity recognition~\cite{inproceedings}. It uses four temporal convolutional blocks with kernel size $5\times 1$, 20 filters, ReLU activations, and batch normalization, followed by channel interaction, channel fusion, temporal interaction, and temporal fusion modules. Temporal interaction is implemented with a 40-unit LSTM, and the fused 40-dimensional representation is mapped to a scalar regression output.

%% file: sections/appendix/foundation_model_transfer_details.tex
For the pretrained/foundation-model transfer experiments, each compatible modality was encoded independently into one fixed-length embedding per supervised window. The encoders were used as frozen feature extractors. The concatenated-embedding baseline used a strict intersection across embedding streams: a supervised window was included only if every modality-specific embedding in Table~\ref{tab:foundation-embedding-details} was available for that same window. This removes windows with missing modality embeddings before model fitting, so the concatenated baseline evaluates complete multimodal representations rather than mixing different modality subsets across samples.

\begin{table}[t]
\centering
\caption{Pretrained/foundation-model embeddings used in the concatenated embedding baseline.}
\label{tab:foundation-embedding-details}
\small
\begin{tabular}{llrp{0.44\linewidth}}
\toprule
Embedding stream & Encoder & Dim. & Input and preprocessing \\
\midrule
ECG & ECGFounder & 1024 & Polar ECG segments resampled to 500 Hz, window-z-scored, and encoded with the 1-lead ECGFounder feature-returning network. \\
Eye/head & MOMENT & 1024 & Beam screen gaze $x,y$ and head position $x,y,z$ resampled to MOMENT context length 512 and participant-normalized. \\
MS Band EDA & MOMENT & 1024 & Resistance converted to log conductance, $\log(1+1000/\mathrm{kOhm})$, resampled to context length 512. \\
MS Band HR & MOMENT & 1024 & Heart-rate bpm, using locked-quality rows when available, resampled to context length 512. \\
Ring temperature & MOMENT & 1024 & Three raw ring temperature channels, two on the inside of the ring and one on the outside, resampled to context length 512. \\
EEG & NeuroLM & 768 & Muse AF7 and AF8 channels resampled to 200 Hz, filtered, divided by 100, tokenized with 10--20 channel aliases, and pooled over valid EEG tokens. \\
eSense IMU & NormWear & 768 & Six-axis eSense acceleration/gyroscope windows resampled to 65 Hz and participant-normalized. \\
Muse IMU & NormWear & 768 & Muse accelerometer and gyroscope channels resampled to 65 Hz and participant-normalized. \\
Muse PPG & PulsePPG & 512 & Muse left outer 730 nm optical channel resampled to 50 Hz and participant-normalized. \\
Ring PPG & PulsePPG & 512 & Ring green optical channel resampled to 50 Hz and participant-normalized. \\
\bottomrule
\end{tabular}
\end{table}

\paragraph{NeuroLM}
NeuroLM was used for EEG representation learning. The retained Muse EEG channels, AF7 and AF8, were mapped to their standard scalp-location aliases. The extractor applied a 0.1 Hz high-pass filter, 75 Hz low-pass filter, and 60 Hz notch filter before resampling to 200 Hz. NeuroLM tokenization groups EEG into one-second tokens. Values were divided by 100 to match the NeuroLM loader convention. The final hidden states were pooled over valid EEG tokens, producing a 768-dimensional embedding per window.

\paragraph{PulsePPG}
PulsePPG was used for optical PPG streams from both the Muse and ring. The Muse extractor used the left outer 730 nm optical channel, while the ring extractor used the green optical channel. Signals were resampled to 50 Hz and normalized using participant/session-level mean and standard deviation. The PulsePPG ResNet1D encoder produced 512-dimensional embeddings.

\paragraph{ECGFounder}
ECGFounder was used for one-lead Polar H10 ECG. Each window was resampled to 500 Hz, z-scored within the window, and passed through the 1-lead ECGFounder Net1D backbone with \texttt{return\_features}\allowbreak\texttt{=True}. The dense diagnostic head was omitted, and the resulting feature vector was 1024-dimensional.

\paragraph{NormWear}
NormWear was used for inertial sensing. We extracted separate embeddings for Muse IMU and eSense IMU. Each input used six channels: acceleration $x,y,z$ and gyroscope $x,y,z$. Signals were resampled to 65 Hz and participant-normalized. NormWear produced patch-level embeddings of shape $[\mathrm{channels},\mathrm{patches},768]$. We pooled by mean over patches and then by mean over channels, yielding a 768-dimensional embedding.

\paragraph{MOMENT}
MOMENT was used for streams without a more specific available biosignal foundation encoder: Beam eye/head features, MS Band EDA, MS Band HR, and ring temperature. Inputs were resampled to a fixed context length of 512. EDA was represented as log conductance from the MS Band resistance channel; HR used bpm values; eye/head used screen gaze and head-position channels; and ring temperature used the three raw temperature channels. MOMENT embeddings were 1024-dimensional.

\subsubsection{Linear Prediction Head for Concatenated Foundation Embeddings}
The concatenated foundation baseline joined the 10 embeddings in Table~\ref{tab:foundation-embedding-details}, producing an 8448-dimensional vector. Within each fold, a \texttt{StandardScaler} was fit only on training participants. The classifier was a single linear layer trained with AdamW, learning rate $10^{-3}$, weight decay $10^{-2}$, batch size 64, maximum 500 epochs, validation fraction 0.2, and early stopping patience 50. Class weights were computed from training-fold class frequencies.

\subsubsection{Gated Fusion of Foundation Embeddings}
We also evaluated a gated fusion baseline using the same strict-intersection windows, participant-based 4-fold cross validation, and target definitions as the concatenated foundation-embedding baseline. Each modality embedding was first projected to a shared 128-dimensional latent space. For each modality $m$, a scalar gate was learned from the projected modality representation and a separate video-progress context feature. The context feature was
\[
    \mathrm{video\ progress} =
    \mathrm{round}\left(
    \frac{(t_{\mathrm{start}}^{\mathrm{video}} + t_{\mathrm{end}}^{\mathrm{video}})/2}
    {\max(t_{\mathrm{end}}^{\mathrm{video}})}
    , 1\right),
\]
computed within each session/video. The context was used by the gates and by the final classifier, but was not concatenated into each modality embedding.

Let $h_m$ denote the projected embedding for modality $m$ and $g_m\in[0,1]$ its sigmoid gate. The fused representation was a normalized gated average,
\[
    z = \frac{\sum_m g_m h_m}{\sum_m g_m + \epsilon}.
\]
The final classifier received $[z;\mathrm{context}]$. In addition to summary metrics, per-fold metrics, and predictions, the gated model saved per-window gate values, allowing post-hoc inspection of modality weighting. Results for the concatenated and gated foundation-embedding baselines are reported together in Table~\ref{tab:supervised_from_scratch_baselines_results}.

%% file: sections/appendix/llm_baseline_details.tex
\edit{For the LLM-based baseline, we first extracted engineered features from the available physiological and behavioral sensor streams for each probe-aligned window. These features were converted into structured text summaries so that the model received compact descriptions of each sensing stream rather than raw time series.}

\edit{Few-shot examples were selected only from participants outside the held-out testing fold. Candidate examples were retrieved using simple similarity criteria, including available sensor streams and coarse feature patterns. We grouped candidate examples by engagement label and selected at most one example from each label, yielding up to five examples for each test window.}

\edit{The prediction procedure followed a lightweight multi-agent structure inspired by ConSensus~\cite{yoon2026consensus}. For each test window, one LLM agent was instantiated for each available sensor stream. Each agent received the task definition, a short description of its stream, the selected few-shot examples, and only the features from that stream. The stream-level agents produced engagement predictions with brief rationales. We then used semantic, statistical, and hybrid fusion prompts to combine the stream-level outputs while accounting for agreement, disagreement, missing streams, and correlated evidence. The final prediction was taken from the hybrid fusion output.}

\edit{We used GPT-5-mini with zero-temperature decoding. Each window required one call per available stream plus three fusion calls, so this baseline was used as a comparison for few-shot multimodal reasoning rather than as the practical deployment target.}

%% file: sections/appendix/data_collection_system.tex
This section summarizes the data collection system used to record multimodal sensing streams during self-guided instructional video learning. The system combined wearable and camera-based sensors with repeated engagement probes. Its purpose was to collect temporally localized self-report labels while preserving enough physiological and behavioral context before each probe to support engagement estimation.

\subsection{Acquisition Topology}
Data collection used a distributed sensing setup rather than a single acquisition computer. This choice was primarily practical: in pilot testing, a single host machine did not reliably maintain simultaneous Bluetooth connections to all devices at the required sampling rates. In deployment, this limitation could potentially be addressed with dedicated Bluetooth adapters, an integrated acquisition hub, or a purpose-built multi-radio collection system. For the study, distributing devices across multiple collectors provided more stable recordings while preserving a shared timestamp basis for post-hoc synchronization.

The participant watched instructional videos on the main experiment laptop while wearing or using the Muse S Athena headband, Microsoft Band 2, Polar H10, eSense earables, ring, and Beam eye tracker. Muse data were relayed through MindMonitor on an iPhone over Open Sound Control (OSC), a lightweight network protocol for streaming time-varying sensor or control messages between applications. The ring was collected through an Android phone. Microsoft Band data were collected through a custom C\# UWP application because the Band SDK required a Windows/UWP environment. Figure~\ref{fig:data-collection-system} illustrates this acquisition topology.

\input{sections/appendix/fig_data_collection_system}

\subsection{Experiment Player and Labels}
The video player was implemented in PsychoPy. It loaded the instructional videos and scheduled probe timestamps, monitored the current video time, displayed the engagement prompt at each scheduled probe, triggered Microsoft Band haptic feedback, and paused playback until the participant responded.

The engagement log included both video-relative time and wall-clock system time. The video timestamp identifies where the prompt occurred within the instructional material, while the wall-clock timestamp, represented as Unix epoch milliseconds in the raw data, is the cross-modal alignment reference.

\subsection{Sensor Streams and Timestamp Provenance}
The synchronization strategy was software timestamp alignment. Each stream was mapped onto Unix epoch time, then windows were cut using absolute start and end times derived from the engagement log.

Muse S Athena recorded EEG, optical PPG, accelerometer, gyroscope, and marker streams. Muse timestamps were assigned by the laptop receiver on packet arrival, so they reflect the receiver clock and include Bluetooth, phone relay, Wi-Fi, and OSC latency. The raw Muse files do not preserve a separate Muse hardware clock.

Beam Eye Tracker provided webcam-based gaze and head-pose features. The host system timestamp recorded at callback arrival was used for alignment; the Beam SDK timestamp was retained as device provenance. The gaze $x,y$ channels represent the participant's estimated screen gaze location, and head $x,y,z$ represents estimated head position relative to the camera.

Polar H10 provided ECG. The host system timestamp recorded at Bluetooth callback arrival was used for alignment, while the Polar device timestamp was treated as internal device provenance. Because each Bluetooth packet contains multiple ECG samples but only one packet timestamp, preprocessing expands repeated packet timestamps into per-sample timestamps using the ECG sampling rate.

Microsoft Band 2 provided wrist electrodermal activity and heart rate. Its raw files contain both \texttt{BandTime}, derived from the Microsoft Band SDK sensor-reading timestamp, and \texttt{SystemTime}, captured by the receiver on the main experiment laptop. The pipeline uses \texttt{SystemTime} as the canonical alignment timestamp because it is on the same clock as the experiment player. \texttt{BandTime} is retained as provenance for clock or relay offsets.

eSense earables provided six-axis IMU data. The collection code stored both a local sample counter and a wall-clock timestamp; the wall-clock timestamp was used for cross-modal alignment. The ring data were collected as packetized logs with embedded packet timestamps. Each ring sample contained optical, IMU, gyroscope, and temperature channels; the three temperature channels correspond to two sensors on the inside of the ring, left and right, and one sensor on the outside.

\subsection{Synchronization Limitations}
This software synchronization approach allowed heterogeneous devices and sampling rates to be combined without a shared hardware trigger, but it also introduces limitations. Timestamp provenance differs by modality: some timestamps are created at device callback time, some at receiver arrival, and some are embedded in exported packets. Cross-machine alignment depends on clock synchronization among the laptops and phones used for collection. Relay paths, especially Muse through phone/OSC and Microsoft Band through UWP forwarding, can introduce latency. Packetized streams such as Polar ECG and ring data also require within-packet timestamp reconstruction.

Despite these limitations, the raw streams generally covered the experiment interval, supporting event-aligned segmentation. The preprocessing pipeline normalizes timestamp units, sorts each stream by time, reconstructs repeated packet timestamps where needed, and slices windows using the absolute engagement-probe timeline.

%% file: sections/appendix/fig_data_collection_system.tex
\begin{figure}[t]
    \centering
    \includegraphics[width=\linewidth]{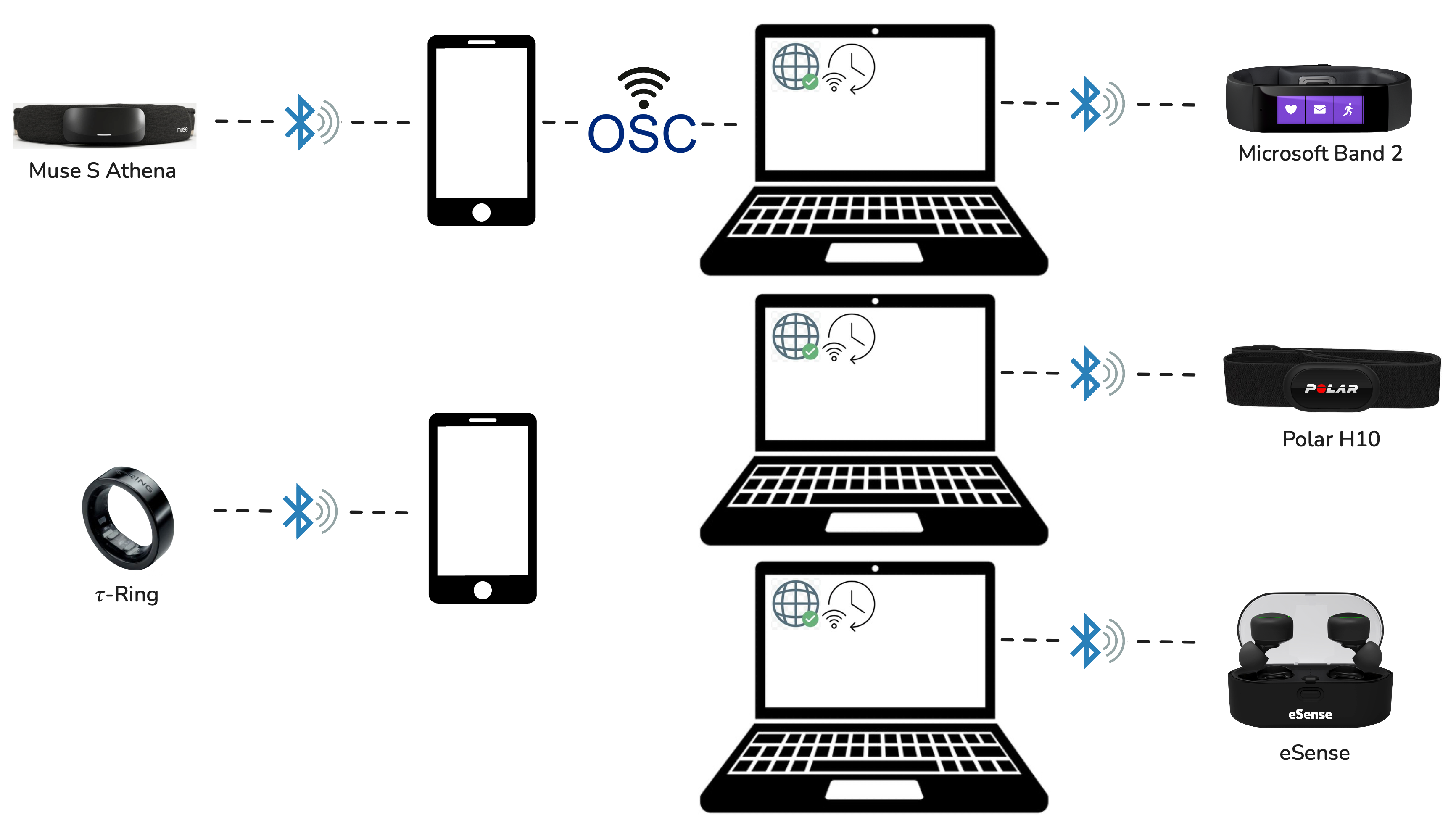}
    \caption{Distributed data collection topology. The Muse headband was relayed through an iPhone using MindMonitor and OSC. The ring was collected through an Android phone. Microsoft Band, Polar H10, and eSense used Bluetooth connections to laptop-based collectors, with the Microsoft Band additionally routed through a custom UWP application on a separate laptop.}
    \label{fig:data-collection-system}
\end{figure}

%% file: sections/appendix/engagement_measure_validation.tex
We conducted a video-level correlation analysis as an external check on the engagement prompt. The prompt asked participants to report how difficult it was to pay attention, with higher values indicating greater attention difficulty. If the prompt captures a meaningful aspect of learning engagement, videos with higher average attention difficulty should tend to show lower learning gains.

For each video/topic, we averaged numeric, non-paused attention-difficulty reports across participants. We also computed normalized learning gain from the pre- and post-video quizzes as
\begin{equation}
    \mathrm{normalized\ gain} = \frac{\mathrm{post} - \mathrm{pre}}{5 - \mathrm{pre}},
\label{eq:normalized-learning-gain}
\end{equation}
where 5 was the maximum quiz score. We then averaged normalized gain across participants for each video/topic and correlated the resulting video-level attention-difficulty and learning-gain values.

Figure~\ref{fig:engagement-learning-gain} shows the relationship across the eight videos/topics. Mean attention difficulty was negatively correlated with normalized learning gain, Pearson $r=-0.75$, $p=0.031$, $n=8$. \edit{This relationship provides preliminary convergent evidence consistent with the intended interpretation of the probe: videos that participants, on average, reported as more difficult to attend to were also associated with lower average normalized learning gains. However, this analysis is based on eight video-level observations and aggregated responses across participants. It therefore does not establish that attention difficulty is equivalent to engagement, validate the measure at the individual level, or demonstrate that the relationship is causal. Rather, it provides an external check showing that the probe varies in an educationally meaningful direction relative to one learning-related outcome.}

\input{sections/appendix/fig_engagement_learning_gain}

%% file: sections/appendix/fig_engagement_learning_gain.tex
\begin{figure}[t]
    \centering
    \includegraphics[width=\linewidth]{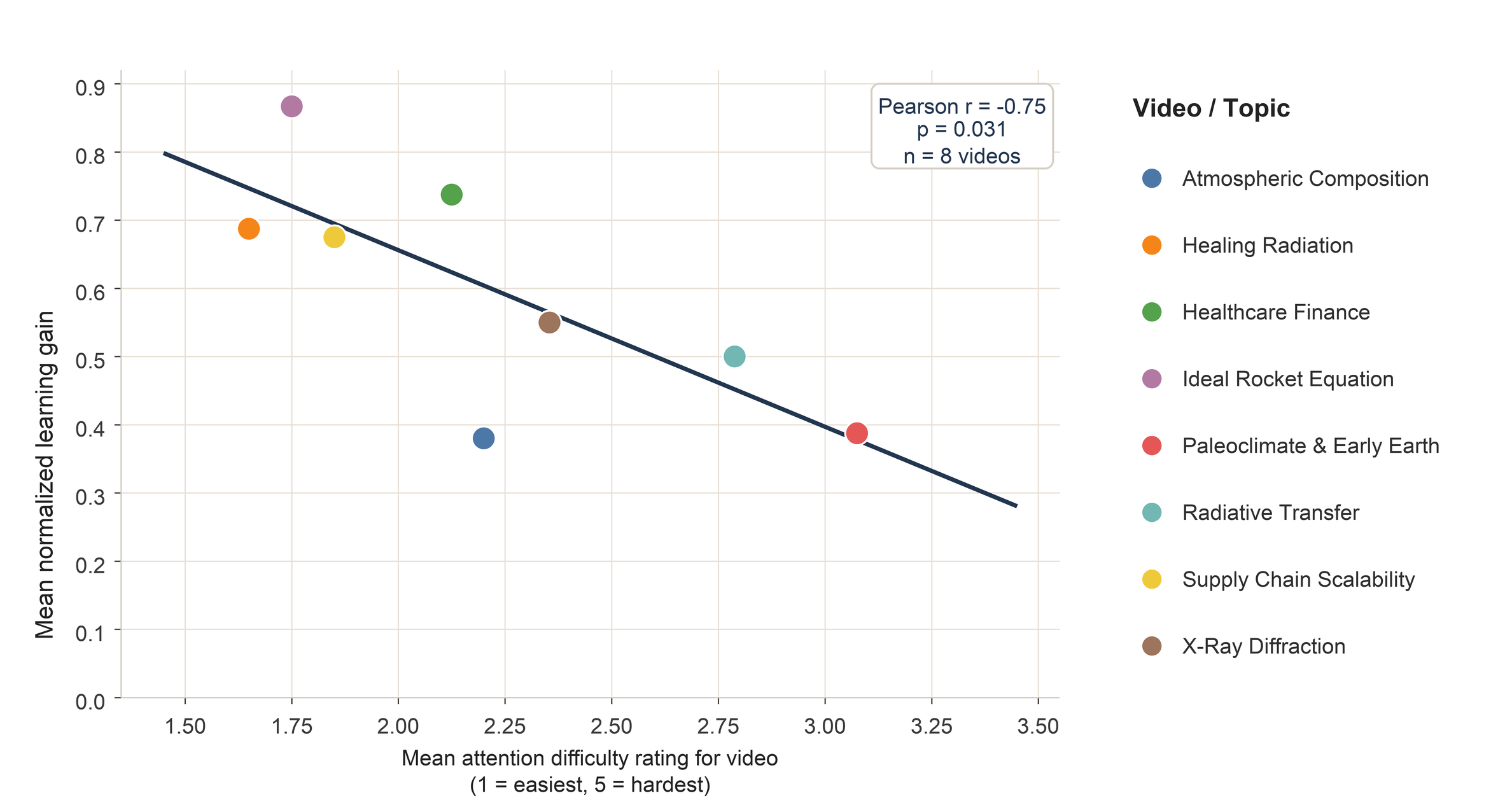}
    \caption{\edit{Video-level relationship between mean attention difficulty and mean normalized learning gain. Each point represents one video/topic averaged across participants. The negative association provides preliminary convergent evidence that the attention-difficulty reports vary in an educationally meaningful direction, Pearson $r=-0.75$, $p=0.031$, $n=8$.}}
    \label{fig:engagement-learning-gain}
\end{figure}

%% file: sections/appendix/demographics.tex
Participants were asked to complete a demographic and background survey prior to the experiment. The questions and possible responses were as follows:

\begin{enumerate}
    \item \textbf{What is your gender?}
    \begin{itemize}
        \item Female
        \item Male
        \item Non-binary
        \item Prefer not to say
    \end{itemize}
    
    \item \textbf{How long do you spend on online learning activities per week?}
    \begin{itemize}
        \item None (0 hours)
        \item 1--3 hours
        \item 4--7 hours
        \item 8--14 hours
        \item More than 15 hours
    \end{itemize}
    
    \item \textbf{Please indicate your major.} \\
    \textit{(Open-ended text response)}
\end{enumerate}

%% file: sections/appendix/videos.tex
Table~\ref{tab:stimulus_videos} details the educational videos used as stimuli during the experiment, spanning four primary disciplines. 

\begin{table}[h]
\centering
\caption{Educational Videos Used in the Study}
\label{tab:stimulus_videos}
\resizebox{\textwidth}{!}{\begin{tabular}{@{}lllll@{}}
\toprule
\textbf{Topic} & \textbf{Subtopic} & \textbf{Video UID} & \textbf{Course / Source} & \textbf{Video Link} \\ \midrule
\textit{Training} & General Training & \texttt{training} & N/A & \url{https://www.youtube.com/watch?v=alhk9eKOLzQ&t=2s} \\ \addlinespace
\multirow{2}{*}{X-Rays} & X-Ray Diffraction & \texttt{xray\_diffraction} & MITx 3.012S.1x & \url{https://www.youtube.com/watch?v=-_P1ki0FRwQ} \\
 & Healing Radiation & \texttt{healing\_radiation} & MITx 22.011x & \url{https://www.youtube.com/watch?v=sxUx27VblDw} \\ \addlinespace
\multirow{2}{*}{Environmental Science} & Paleoclimate & \texttt{paleoclimate} & MITx 12.340x & \url{https://www.youtube.com/watch?v=SWx1pmde63E} \\
 & Radiative Transfer & \texttt{radiative\_transfer} & MITx 12.340x & \url{https://www.youtube.com/watch?v=Afw4fNGEqig} \\ \addlinespace
\multirow{2}{*}{Business} & Supply Chain Scalability & \texttt{supply\_chain\_scalability} & MITx CITE101x & \url{https://www.youtube.com/watch?v=ul3G43yXIFo} \\
 & Healthcare Finance\textsuperscript{*} & \texttt{healthcare} & MITx 15.482x & \makecell[l]{\url{https://www.youtube.com/watch?v=IK6gWreAGOA} \\ \url{https://www.youtube.com/watch?v=MS9WRP2gVo4}} \\ \addlinespace
\multirow{2}{*}{Aerospace Engineering} & Ideal Rocket Equation & \texttt{ideal\_rocket\_equation} & MITx 16.00x & \url{https://www.youtube.com/watch?v=Ddp2HQROuaA} \\
 & Atmospheric Pressure \& Composition & \texttt{atmosphere\_pressure\_composition} & MITx 16.00x & \url{https://www.youtube.com/watch?v=Kc7whUqjWu8} \\ \bottomrule
\multicolumn{5}{l}{\small \textsuperscript{*}Presented continuously as a single stitched video during the experiment to form a 10-minute segment.}
\end{tabular}}
\end{table}

%% file: sections/appendix/prompt.tex
Prior to the actual experiment, participants were presented with the following \edit{instructions explaining the engagement-probe response scale}:

\begin{quote}
\textbf{Prompt:} \\
How difficult was it to pay attention during the last minute of the lecture? (1-5)

\begin{itemize}
    \item[\textbf{1:}] My attention is fused with the lecture; it is completely automatic and effortless.
    \item[\textbf{2}] 
    \item[\textbf{3}] 
    \item[\textbf{4}] 
    \item[\textbf{5:}] I am forcing my attention. It feels like a heavy, conscious struggle to keep up with the lesson.
    \item[\textbf{X:}] External Distraction/Interruption. Use this if something outside the lecture forced your attention away entirely.
\end{itemize}

When it is time to report, the band will vibrate, the video will pause, and the screen will show ``Report engagement now.'' Press 1, 2, 3, 4, 5, or X to report. The video will resume after you respond.

Press SPACE to continue.
\end{quote}

\edit{Only the endpoints of the five-point scale were verbally anchored; values 2--4 represented intermediate response levels, and no additional verbal descriptions were displayed for them. During the study, the full item and anchors were not repeated at every probe. Instead, the video paused approximately every minute, the wearable band vibrated, and the screen displayed only the abbreviated cue \textbf{``Report engagement now.''} Participants then pressed the corresponding key (1, 2, 3, 4, 5, or X), after which the video immediately resumed. Thus, ``Report engagement now'' referred back to the engagement-probe scale that participants had learned and practiced; it was not a separate measurement item.}

%% file: sections/appendix/quizzes.tex
The comprehension quiz instruments are included in the released \projectname{} dataset on Figshare together with the synchronized sensor data, quiz responses, study metadata, and other study materials. The dataset version includes a plaintext file listing all quiz questions and answer choices, with correct answer choices marked.

%% file: sections/appendix/cross_validation_explanation.tex
\begin{figure}[t]
    \centering
    \includegraphics[width=\linewidth]{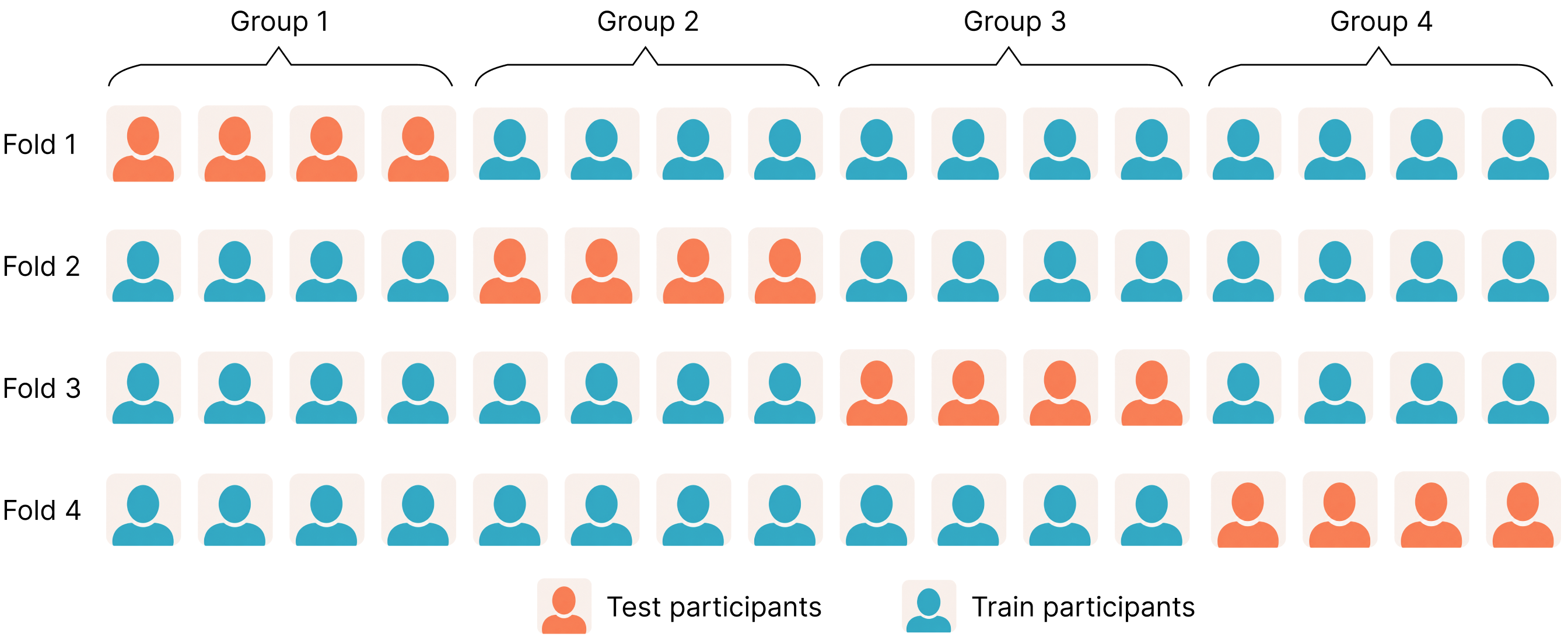}
    \caption{\edit{Participant-grouped 4-fold cross-validation protocol.}}
    \label{fig:participant-grouped-cv}
\end{figure}

\edit{To evaluate generalization to held-out learners within this cohort, we used participant-grouped 4-fold cross-validation. Before model evaluation, the 16 participants were partitioned into four fixed, disjoint groups: Group 1 contained P1, P2, P3, and P10; Group 2 contained P4, P5, P7, and P9; Group 3 contained P6, P8, P11, and P12; and Group 4 contained P13, P14, P15, and P16. Each group contained samples spanning all five response ratings, and the same groups were used across all evaluated methods. As shown in Figure~\ref{fig:participant-grouped-cv}, each fold held out one complete participant group for testing and used the remaining three groups for training. All windows from a participant remained in that participant's assigned group; therefore, no participant contributed windows to both the training and test sets within a fold. Preprocessing parameters, normalization statistics, model training, hyperparameter selection, and checkpoint selection were determined using only the training-side data, while the held-out group was used only for final evaluation. This procedure prevents participant-level data leakage between training and testing.}

\edit{We used participant-grouped 4-fold cross-validation rather than leave-one-subject-out validation because individual participants did not always provide samples covering the full rating range. Holding out only one participant could therefore produce test folds with missing or highly imbalanced labels, resulting in highly variable or poorly defined fold-level metrics, particularly binary Macro-F1 when a held-out participant lacks one of the binary classes. Holding out groups of participants preserves participant independence while ensuring that every test fold contains all five ratings and both binary classes. We computed the metrics separately for each held-out fold and report the final performance as the mean and standard deviation across the four folds.}

%% file: sections/appendix/deployment_footprint.tex
\edit{Table~\ref{tab:deployment-footprint} reports practical deployment costs for the evaluated benchmark models. Experiments were run on the Zatopek server, which has two AMD EPYC 7313 16-core CPUs, 64 logical CPU threads, and 251 GiB RAM; inference timing was measured per engagement window with batch size 1 on a single NVIDIA RTX A6000 GPU. Model size is computed as the FP32 parameter or fitted-state footprint using 4 bytes per parameter-equivalent value, and training time is reported as the mean per-fold time across the fixed four-fold evaluation. For foundation-transfer baselines, size includes the downstream prediction head and one copy of each frozen foundation encoder needed to extract embeddings at deployment; training and inference times are measured after embeddings have been precomputed and therefore exclude frozen-encoder extraction.}

\begin{table*}[t]
    \centering
    \caption{\edit{Model footprint and runtime measurements. Train time is the mean per-fold time across the fixed four-fold evaluation. Inference is measured with batch size 1; values are mean ± standard deviation across the 699 engagement windows where applicable.}}
    \label{tab:deployment-footprint}
    \scriptsize
    \begin{tabular}{@{}lccc@{}}
        \toprule
        \edit{Model} & \edit{Size (MB)} & \edit{Train Time/Fold} & \edit{Infer./Window} \\
        \midrule
        \multicolumn{4}{@{}l}{\edit{\textit{Sensor-Free Baselines}}} \\
        \midrule
        \edit{Mean} & \edit{0.00} & \edit{0.14 ms} & \edit{0.00 ms $\pm$ 0.00 ms} \\
        \edit{Mode} & \edit{0.00} & \edit{0.14 ms} & \edit{0.00 ms $\pm$ 0.00 ms} \\
        \edit{Random Distribution} & \edit{0.00} & \edit{0.14 ms} & \edit{0.00 ms $\pm$ 0.00 ms} \\
        \edit{Peer Window Mean ($\pm 5$ s)} & \edit{0.00} & \edit{0.14 ms} & \edit{0.00 ms $\pm$ 0.00 ms} \\
        \midrule
        \multicolumn{4}{@{}l}{\edit{\textit{ML Statistical Regressors}}} \\
        \midrule
        \edit{Random Forest} & \edit{5.50} & \edit{755.3 ms} & \edit{91.2 ms $\pm$ 21.3 ms} \\
        \edit{Linear Regression} & \edit{0.01} & \edit{29.5 ms} & \edit{1.3 ms $\pm$ 0.01 ms} \\
        \edit{LightGBM} & \edit{0.82} & \edit{11.4 s} & \edit{2.0 ms $\pm$ 0.85 ms} \\
        \edit{SVM} & \edit{0.86} & \edit{35.9 ms} & \edit{1.4 ms $\pm$ 0.02 ms} \\
        \midrule
        \multicolumn{4}{@{}l}{\edit{\textit{Sequence-Based DL Baselines}}} \\
        \midrule
        \edit{DeepConvLSTM} & \edit{4.70} & \edit{24.9 s} & \edit{10.5 ms $\pm$ 0.28 ms} \\
        \edit{DeepConvLSTM with Self-Attention} & \edit{0.03} & \edit{4.1 s} & \edit{1.3 ms $\pm$ 0.06 ms} \\
        \edit{TinyHAR} & \edit{0.18} & \edit{3.5 min} & \edit{104.6 ms $\pm$ 1.1 ms} \\
        \midrule
        \multicolumn{4}{@{}l}{\edit{\textit{Foundation-Model Transfer Baselines}}} \\
        \midrule
        \edit{Concat Embeddings + Linear Head} & \edit{5,230.38} & \edit{133.5 ms} & \edit{0.06 ms $\pm$ 0.01 ms} \\
        \edit{Gated Fusion} & \edit{5,236.40} & \edit{1.8 s} & \edit{2.0 ms $\pm$ 0.03 ms} \\
        \midrule
        \multicolumn{4}{@{}l}{\edit{\textit{LLM-Based Reasoning Baseline}}} \\
        \midrule
        \edit{LLM-Based Few-shot} & \edit{0.00} & \edit{0.00} & \edit{71.52 s $\pm$ 1.66 s } \\
        \midrule
        \edit{Modality-Aware Model} & \edit{3.50} & \edit{1.0 min} & \edit{14.0 ms $\pm$ 0.72 ms} \\
        \bottomrule
    \end{tabular}
\end{table*}

\edit{Training the larger models may still be most practical on a server-class GPU, especially when rerunning cross-validation or hyperparameter sweeps. Deployment is less demanding for the task-specific non-LLM models: after training, they require at most 6.2 MB of fitted model state and use single-window inference times below 105 ms on the A6000 GPU, with most models below 15 ms. The foundation-transfer baselines require substantially more memory because deployment must include the frozen encoders used to produce their embeddings. These small memory and latency requirements for the task-specific models suggest that inference should be feasible on lower-tier GPUs or CPUs, although we do not claim identical latency on hardware we did not benchmark directly.}

\edit{For environmental cost, we estimate conservative upper bounds from measured runtime and the A6000's 300 W maximum board power~\cite{nvidiaRTXA6000Spec}, using $E_{\mathrm{kWh}}=(300 \times t_{\mathrm{s}} \times N)/(3.6 \times 10^6)$. Using the UNU-INWEH report's 2030 data-center footprint totals as intensity factors, $399$ MtCO$_2$e, $9.3$ trillion L water, and $14{,}500$ km$^2$ land over $945$ TWh~\cite{aczel2026environmentalAI}, the modality-aware model's per-fold training cost ($t=60$ s, $N=1$) is estimated at $0.0050$ kWh, or $2.11$ gCO$_2$e, $49.2$ mL water, and $7.67\times10^{-5}$ m$^2$ land. The corresponding four-fold rerun is approximately $0.020$ kWh, or $8.44$ gCO$_2$e, $196.8$ mL water, and $3.07\times10^{-4}$ m$^2$ land. Its inference footprint for one full pass over the dataset ($t=0.014$ s, $N=699$) is much smaller: $8.16\times10^{-4}$ kWh, or $0.34$ gCO$_2$e, $8.0$ mL water, and $1.25\times10^{-5}$ m$^2$ land. For context, the slowest measured neural model, TinyHAR, has an upper-bound per-fold training footprint of $7.39$ gCO$_2$e, $172.2$ mL water, and $2.68\times10^{-4}$ m$^2$ land, and a full-dataset inference footprint of $2.57$ gCO$_2$e, $60.0$ mL water, and $9.35\times10^{-5}$ m$^2$ land. Hyperparameter sweeps scale approximately with the number of runs. These estimates exclude device manufacturing, sensor operation, and foundation-model pretraining, and likely overestimate runtime electricity because they assume maximum GPU power throughout training and inference.}

\edit{Monetary cost depends mainly on sensing hardware and whether inference uses an external LLM. The trained non-LLM models do not require paid API calls at deployment. In contrast, our few-shot LLM baseline used GPT-5-mini and makes paid model calls, costing approximately \$0.10 per engagement-window prediction in our billing estimate, or about \$69.90 for all 699 windows; for this reason, we treat it as a reference baseline rather than the practical deployment target. For sensing hardware, public commercial prices at the time of writing were approximately \$474.99 for Muse S Athena~\cite{museSAthenaPrice}, \$104.95 for Polar H10~\cite{polarH10Price}, \$29.99 for Beam Eye Tracker software~\cite{beamEyeTrackerPrice}, and \$39.99 for the Logitech Brio webcam used with Beam~\cite{logitechBrio100Price}. Microsoft Band 2 is discontinued, but launched at \$249.99~\cite{microsoftBand2Price}. The $\tau$-Ring and eSense earable were research or open-platform devices rather than stable retail products~\cite{tang2025ring,kawsar2018esense}, so we do not include them in a reproducible commercial total. The reduced-device analyses in Section~\ref{sec:modality_analysis} are intended to support lower-cost deployments that use fewer sensors.}

%% file: bibs/ref.bib
@article{kulsoom2022review,
  title={A review of machine learning-based human activity recognition for diverse applications},
  author={Kulsoom, Farzana and Narejo, Sanam and Mehmood, Zahid and Chaudhry, Hassan Nazeer and Butt, Ayesha and Bashir, Ali Kashif},
  journal={Neural Computing and Applications},
  volume={34},
  number={21},
  pages={18289--18324},
  year={2022},
  publisher={Springer}
}

@inproceedings{inproceedings,
author = {Mohammed, Sara and Elbasiony, Reda and Gomaa, Walid},
year = {2018},
month = {01},
pages = {504-511},
title = {An LSTM-based Descriptor for Human Activities Recognition using IMU Sensors},
doi = {10.5220/0006902405040511}
}

@ARTICLE{singh2021DeepConvLSTM,
  author={Singh, Satya P. and Sharma, Madan Kumar and Lay-Ekuakille, Aimé and Gangwar, Deepak and Gupta, Sukrit},
  journal={IEEE Sensors Journal}, 
  title={Deep ConvLSTM With Self-Attention for Human Activity Decoding Using Wearable Sensors}, 
  year={2021},
  volume={21},
  number={6},
  pages={8575-8582},
  doi={10.1109/JSEN.2020.3045135}}

@article{fredricks2004school,
  title={School engagement: Potential of the concept, state of the evidence},
  author={Fredricks, Jennifer A and Blumenfeld, Phyllis C and Paris, Alison H},
  journal={Review of educational research},
  volume={74},
  number={1},
  pages={59--109},
  year={2004},
  publisher={Sage Publications Sage CA: Thousand Oaks, CA}
}

@inproceedings{guo2014video,
  title={How video production affects student engagement: An empirical study of MOOC videos},
  author={Guo, Philip J and Kim, Juho and Rubin, Rob},
  booktitle={Proceedings of the first ACM conference on Learning@ scale conference},
  pages={41--50},
  year={2014}
}

@article{miller2013comparison,
  title={A comparison of traditional and engaging lecture methods in a large, professional-level course},
  author={Miller, Cynthia J and McNear, Jacquee and Metz, Michael J},
  journal={Advances in physiology education},
  volume={37},
  number={4},
  pages={347--355},
  year={2013},
  publisher={American Physiological Society Bethesda, MD}
}

@article{krathwohl2002revision,
  title={A revision of Bloom's taxonomy: An overview},
  author={Krathwohl, David R},
  journal={Theory into practice},
  volume={41},
  number={4},
  pages={212--218},
  year={2002},
  publisher={Taylor \& Francis}
}

@incollection{blumenfeld2006motivation,
  author    = {Blumenfeld, Phyllis C. and Kempler, Tali M. and Krajcik, Joseph S.},
  title     = {Motivation and Cognitive Engagement in Learning Environments},
  booktitle = {The Cambridge Handbook of the Learning Sciences},
  editor    = {Sawyer, R. Keith},
  pages     = {475--488},
  year      = {2006},
  publisher = {Cambridge University Press}
}

@article{chi2014icap,
  title={The ICAP framework: Linking cognitive engagement to active learning outcomes},
  author={Chi, Michelene TH and Wylie, Ruth},
  journal={Educational psychologist},
  volume={49},
  number={4},
  pages={219--243},
  year={2014},
  publisher={Taylor \& Francis}
}

@article{pintrich1990motivational,
  title={Motivational and self-regulated learning components of classroom academic performance.},
  author={Pintrich, Paul R and De Groot, Elisabeth V},
  journal={Journal of educational psychology},
  volume={82},
  number={1},
  pages={33},
  year={1990},
  publisher={American Psychological Association}
}

@article{conrad2021measuring,
  title={Measuring mind wandering during online lectures assessed with EEG},
  author={Conrad, Colin and Newman, Aaron},
  journal={Frontiers in Human Neuroscience},
  volume={15},
  pages={697532},
  year={2021},
  publisher={Frontiers Media SA}
}

@inproceedings{kizilcec2013deconstructing,
  title={Deconstructing disengagement: analyzing learner subpopulations in massive open online courses},
  author={Kizilcec, Ren{\'e} F and Piech, Chris and Schneider, Emily},
  booktitle={Proceedings of the third international conference on learning analytics and knowledge},
  pages={170--179},
  year={2013}
}

@article{henrie2015measuring,
  title={Measuring student engagement in technology-mediated learning: A review},
  author={Henrie, Curtis R and Halverson, Lisa R and Graham, Charles R},
  journal={Computers \& Education},
  volume={90},
  pages={36--53},
  year={2015},
  publisher={Elsevier}
}

@article{smallwood2015science,
  title={The science of mind wandering: Empirically navigating the stream of consciousness},
  author={Smallwood, Jonathan and Schooler, Jonathan W},
  journal={Annual review of psychology},
  volume={66},
  number={1},
  pages={487--518},
  year={2015},
  publisher={Annual Reviews}
}

@article{gao2020n,
  title={n-gage: Predicting in-class emotional, behavioural and cognitive engagement in the wild},
  author={Gao, Nan and Shao, Wei and Rahaman, Mohammad Saiedur and Salim, Flora D},
  journal={Proceedings of the ACM on Interactive, Mobile, Wearable and Ubiquitous Technologies},
  volume={4},
  number={3},
  pages={1--26},
  year={2020},
  publisher={ACM New York, NY, USA}
}

@article{horvers2021detecting,
  title={Detecting emotions through electrodermal activity in learning contexts: A systematic review},
  author={Horvers, Anne and Tombeng, Natasha and Bosse, Tibor and Lazonder, Ard W and Molenaar, Inge},
  journal={Sensors},
  volume={21},
  number={23},
  pages={7869},
  year={2021},
  publisher={MDPI}
}

@article{bustos2022wearables,
  title={Wearables for engagement detection in learning environments: A review},
  author={Bustos-Lopez, Maritza and Cruz-Ramirez, Nicandro and Guerra-Hernandez, Alejandro and S{\'a}nchez-Morales, Laura Nely and Cruz-Ramos, Nancy Aracely and Alor-Hernandez, Giner},
  journal={Biosensors},
  volume={12},
  number={7},
  pages={509},
  year={2022},
  publisher={MDPI}
}

@article{pope1995biocybernetic,
  title={Biocybernetic system evaluates indices of operator engagement in automated task},
  author={Pope, Alan T and Bogart, Edward H and Bartolome, Debbie S},
  journal={Biological psychology},
  volume={40},
  number={1-2},
  pages={187--195},
  year={1995},
  publisher={Elsevier}
}

@article{disalvo2022reading,
  title={Reading the room: Automated, momentary assessment of student engagement in the classroom: Are we there yet?},
  author={DiSalvo, Betsy and Bandaru, Dheeraj and Wang, Qiaosi and Li, Hong and Pl{\"o}tz, Thomas},
  journal={Proceedings of the ACM on Interactive, Mobile, Wearable and Ubiquitous Technologies},
  volume={6},
  number={3},
  pages={1--26},
  year={2022},
  publisher={ACM New York, NY, USA}
}

@article{whitehill2014faces,
  title={The faces of engagement: Automatic recognition of student engagementfrom facial expressions},
  author={Whitehill, Jacob and Serpell, Zewelanji and Lin, Yi-Ching and Foster, Aysha and Movellan, Javier R},
  journal={IEEE Transactions on Affective Computing},
  volume={5},
  number={1},
  pages={86--98},
  year={2014},
  publisher={IEEE}
}

@article{monkaresi2016automated,
  title={Automated detection of engagement using video-based estimation of facial expressions and heart rate},
  author={Monkaresi, Hamed and Bosch, Nigel and Calvo, Rafael A and D'Mello, Sidney K},
  journal={IEEE Transactions on Affective Computing},
  volume={8},
  number={1},
  pages={15--28},
  year={2016},
  publisher={IEEE}
}

@article{di2018unobtrusive,
  title={Unobtrusive assessment of students' emotional engagement during lectures using electrodermal activity sensors},
  author={Di Lascio, Elena and Gashi, Shkurta and Santini, Silvia},
  journal={Proceedings of the ACM on Interactive, Mobile, Wearable and Ubiquitous Technologies},
  volume={2},
  number={3},
  pages={1--21},
  year={2018},
  publisher={ACM New York, NY, USA}
}

@article{d2012dynamics,
  title={Dynamics of affective states during complex learning},
  author={D’Mello, Sidney and Graesser, Art},
  journal={Learning and Instruction},
  volume={22},
  number={2},
  pages={145--157},
  year={2012},
  publisher={Elsevier}
}

@article{calvo2010affect,
  title={Affect detection: An interdisciplinary review of models, methods, and their applications},
  author={Calvo, Rafael A and D'Mello, Sidney},
  journal={IEEE Transactions on affective computing},
  volume={1},
  number={1},
  pages={18--37},
  year={2010},
  publisher={IEEE}
}

@article{berntson1997heart,
  title={Heart rate variability: origins, methods, and interpretive caveats},
  author={Berntson, Gary G and Thomas Bigger Jr, J and Eckberg, Dwain L and Grossman, Paul and Kaufmann, Peter G and Malik, Marek and Nagaraja, Haikady N and Porges, Stephen W and Saul, J Philip and Stone, Peter H and others},
  journal={Psychophysiology},
  volume={34},
  number={6},
  pages={623--648},
  year={1997},
  publisher={Wiley Online Library}
}

@book{boucsein2012electrodermal,
  title={Electrodermal activity},
  author={Boucsein, Wolfram},
  year={2012},
  publisher={Springer science \& business media}
}

@article{appleton2006measuring,
  title={Measuring cognitive and psychological engagement: Validation of the Student Engagement Instrument},
  author={Appleton, James J and Christenson, Sandra L and Kim, Dongjin and Reschly, Amy L},
  journal={Journal of school psychology},
  volume={44},
  number={5},
  pages={427--445},
  year={2006},
  publisher={Elsevier}
}

@article{dejonckheere2022assessing,
  title={Assessing the reliability of single-item momentary affective measurements in experience sampling.},
  author={Dejonckheere, Egon and Demeyer, Febe and Geusens, Birte and Piot, Maarten and Tuerlinckx, Francis and Verdonck, Stijn and Mestdagh, Merijn},
  journal={Psychological assessment},
  volume={34},
  number={12},
  pages={1138},
  year={2022},
  publisher={American Psychological Association}
}

@article{song2023examining,
  title={Examining the concurrent and predictive validity of single items in ecological momentary assessments},
  author={Song, Jiyoung and Howe, Esther and Oltmanns, Joshua R and Fisher, Aaron J},
  journal={Assessment},
  volume={30},
  number={5},
  pages={1662--1671},
  year={2023},
  publisher={Sage Publications Sage CA: Los Angeles, CA}
}

@incollection{shernoff2014student,
  title={Student engagement in high school classrooms from the perspective of flow theory},
  author={Shernoff, David J and Csikszentmihalyi, Mihaly and Schneider, Barbara and Shernoff, Elisa Steele},
  booktitle={Applications of flow in human development and education: The collected works of Mihaly Csikszentmihalyi},
  pages={475--494},
  year={2014},
  publisher={Springer}
}

@article{shernof2017student,
  title={Student engagement as a general factor of classroom experience: Associations with student practices and educational outcomes in a university gateway course},
  author={Shernof, David J and Ruzek, Erik A and Sannella, Alexander J and Schorr, Roberta Y and Sanchez-Wall, Lina and Bressler, Denise M},
  journal={Frontiers in psychology},
  volume={8},
  pages={994},
  year={2017},
  publisher={Frontiers Media SA}
}

@article{de2024measuring,
  title={Measuring student engagement in lessons using an experience sampling methodology: The development and validation of the dynamic engagement with learning questionnaire},
  author={De Weerdt, Dries and Simons, Mathea and Struyf, Elke},
  journal={Journal of Psychoeducational Assessment},
  volume={42},
  number={5},
  pages={527--539},
  year={2024},
  publisher={SAGE Publications Sage CA: Los Angeles, CA}
}

@article{martin2020factors,
  title={What factors influence students’ real-time motivation and engagement? An experience sampling study of high school students using mobile technology},
  author={Martin, Andrew J and Mansour, Marianne and Malmberg, Lars-Erik},
  journal={Educational Psychology},
  volume={40},
  number={9},
  pages={1113--1135},
  year={2020},
  publisher={Taylor \& Francis}
}

@article{xie2019examining,
  title={Examining engagement in context using experience-sampling method with mobile technology},
  author={Xie, Kui and Heddy, Benjamin C and Vongkulluksn, Vanessa W},
  journal={Contemporary Educational Psychology},
  volume={59},
  pages={101788},
  year={2019},
  publisher={Elsevier}
}

@article{manwaring2017investigating,
  title={Investigating student engagement in blended learning settings using experience sampling and structural equation modeling},
  author={Manwaring, Kristine C and Larsen, Ross and Graham, Charles R and Henrie, Curtis R and Halverson, Lisa R},
  journal={The Internet and Higher Education},
  volume={35},
  pages={21--33},
  year={2017},
  publisher={Elsevier}
}

@article{apicella2022eeg,
  title={EEG-based measurement system for monitoring student engagement in learning 4.0},
  author={Apicella, Andrea and Arpaia, Pasquale and Frosolone, Mirco and Improta, Giovanni and Moccaldi, Nicola and Pollastro, Andrea},
  journal={Scientific Reports},
  volume={12},
  number={1},
  pages={5857},
  year={2022},
  publisher={Nature Publishing Group UK London}
}

@article{gao2022individual,
  title={Individual and group-wise classroom seating experience: Effects on student engagement in different courses},
  author={Gao, Nan and Rahaman, Mohammad Saiedur and Shao, Wei and Ji, Kaixin and Salim, Flora D},
  journal={Proceedings of the ACM on Interactive, Mobile, Wearable and Ubiquitous Technologies},
  volume={6},
  number={3},
  pages={1--23},
  year={2022},
  publisher={ACM New York, NY, USA}
}

@article{schroeder2023scoping,
  title={A scoping review of wrist-worn wearables in education},
  author={Schroeder, Noah L and Romine, William L and Kemp, Sidney E},
  journal={Computers and Education Open},
  volume={5},
  pages={100154},
  year={2023},
  publisher={Elsevier}
}

@article{savchenko2022classifying,
  title={Classifying emotions and engagement in online learning based on a single facial expression recognition neural network},
  author={Savchenko, Andrey V and Savchenko, Lyudmila V and Makarov, Ilya},
  journal={IEEE transactions on affective computing},
  volume={13},
  number={4},
  pages={2132--2143},
  year={2022},
  publisher={IEEE}
}

@article{kahu2013framing,
  title={Framing student engagement in higher education},
  author={Kahu, Ella R},
  journal={Studies in higher education},
  volume={38},
  number={5},
  pages={758--773},
  year={2013},
  publisher={Taylor \& Francis}
}

@article{wong2022student,
  title={Student engagement: Current state of the construct, conceptual refinement, and future research directions},
  author={Wong, Zi Yang and Liem, Gregory Arief D},
  journal={Educational Psychology Review},
  volume={34},
  number={1},
  pages={107--138},
  year={2022},
  publisher={Springer}
}

@article{szpunar2013interpolated,
  title={Interpolated memory tests reduce mind wandering and improve learning of online lectures},
  author={Szpunar, Karl K and Khan, Novall Y and Schacter, Daniel L},
  journal={Proceedings of the National Academy of Sciences},
  volume={110},
  number={16},
  pages={6313--6317},
  year={2013},
  publisher={National Academy of Sciences}
}

@article{kahu2018student,
  title={Student engagement in the educational interface: Understanding the mechanisms of student success},
  author={Kahu, Ella R and Nelson, Karen},
  journal={Higher education research \& development},
  volume={37},
  number={1},
  pages={58--71},
  year={2018},
  publisher={Taylor \& Francis}
}

@article{noetel2021video,
  title={Video improves learning in higher education: A systematic review},
  author={Noetel, Michael and Griffith, Shantell and Delaney, Oscar and Sanders, Taren and Parker, Philip and del Pozo Cruz, Borja and Lonsdale, Chris},
  journal={Review of educational research},
  volume={91},
  number={2},
  pages={204--236},
  year={2021},
  publisher={Sage Publications Sage CA: Los Angeles, CA}
}

@article{kuhlmann2024students,
  title={Students’ active cognitive engagement with instructional videos predicts STEM learning},
  author={Kuhlmann, Shelbi L and Plumley, Robert and Evans, Zoe and Bernacki, Matthew L and Greene, Jeffrey A and Hogan, Kelly A and Berro, Michael and Gates, Kathleen and Panter, Abigail},
  journal={Computers \& Education},
  volume={216},
  pages={105050},
  year={2024},
  publisher={Elsevier}
}

@article{anders2024associations,
  title={Associations between mind wandering, viewer interactions, and the meaningful structure of educational videos},
  author={Anders, Gerrit and Buder, J{\"u}rgen and Merkt, Martin and Egger, Etienne and Huff, Markus},
  journal={Computers \& Education},
  volume={212},
  pages={104996},
  year={2024},
  publisher={Elsevier}
}

@inproceedings{kim2014understanding,
  title={Understanding in-video dropouts and interaction peaks inonline lecture videos},
  author={Kim, Juho and Guo, Philip J and Seaton, Daniel T and Mitros, Piotr and Gajos, Krzysztof Z and Miller, Robert C},
  booktitle={Proceedings of the first ACM conference on Learning@ scale conference},
  pages={31--40},
  year={2014}
}

@article{csikszentmihalyi1987validity,
  title={Validity and reliability of the experience-sampling method},
  author={Csikszentmihalyi, Mihaly and Larson, Reed},
  journal={The Journal of nervous and mental disease},
  volume={175},
  number={9},
  pages={526--536},
  year={1987},
  publisher={LWW}
}

@article{shiffman2008ecological,
  title={Ecological momentary assessment},
  author={Shiffman, Saul and Stone, Arthur A and Hufford, Michael R},
  journal={Annu. Rev. Clin. Psychol.},
  volume={4},
  number={1},
  pages={1--32},
  year={2008},
  publisher={Annual Reviews}
}

@article{critchley2002electrodermal,
  title={Electrodermal responses: what happens in the brain},
  author={Critchley, Hugo D},
  journal={The Neuroscientist},
  volume={8},
  number={2},
  pages={132--142},
  year={2002},
  publisher={SAGE Publications Sage CA: Los Angeles, CA}
}

@article{cinaz2013monitoring,
  title={Monitoring of mental workload levels during an everyday life office-work scenario},
  author={Cinaz, Burcu and Arnrich, Bert and La Marca, Roberto and Tr{\"o}ster, Gerhard},
  journal={Personal and ubiquitous computing},
  volume={17},
  number={2},
  pages={229--239},
  year={2013},
  publisher={Springer}
}

@article{beh2021robust,
  title={Robust PPG-based mental workload assessment system using wearable devices},
  author={Beh, Win-Ken and Wu, Yi-Hsuan and Wu, An-Yeu},
  journal={IEEE Journal of Biomedical and Health Informatics},
  volume={27},
  number={5},
  pages={2323--2333},
  year={2021},
  publisher={IEEE}
}

@article{mach2022assessing,
  title={Assessing mental workload with wearable devices--Reliability and applicability of heart rate and motion measurements},
  author={Mach, Sebastian and Storozynski, Pamela and Halama, Josephine and Krems, Josef F},
  journal={Applied ergonomics},
  volume={105},
  pages={103855},
  year={2022},
  publisher={Elsevier}
}

@article{dan2017real,
  title={Real time EEG based measurements of cognitive load indicates mental states during learning},
  author={Dan, Alex and Reiner, Miriam and others},
  journal={Journal of Educational Data Mining},
  volume={9},
  number={2},
  pages={31--44},
  year={2017}
}

@article{sarailoo2022assessment,
  title={Assessment of instantaneous cognitive load imposed by educational multimedia using electroencephalography signals},
  author={Sarailoo, Reza and Latifzadeh, Kayhan and Amiri, S Hamid and Bosaghzadeh, Alireza and Ebrahimpour, Reza},
  journal={Frontiers in neuroscience},
  volume={16},
  pages={744737},
  year={2022},
  publisher={Frontiers Media SA}
}

@article{pei2025design,
  title={Design and validation of an electroencephalogram-supported approach to tracking real-time cognitive load variations for adaptive video-based learning},
  author={Pei, Leisi and Jong, Morris Siu-Yung and Shang, Junjie and Ouyang, Guang},
  journal={British Journal of Educational Technology},
  volume={56},
  number={4},
  pages={1553--1572},
  year={2025},
  publisher={Wiley Online Library}
}

@article{hutt2017gaze,
  title={Gaze-Based Detection of Mind Wandering during Lecture Viewing.},
  author={Hutt, Stephen and Hardey, Jessica and Bixler, Robert and Stewart, Angela and Risko, Evan and D'Mello, Sidney K},
  journal={International Educational Data Mining Society},
  year={2017},
  publisher={ERIC}
}

@article{hutt2019automated,
  title={Automated gaze-based mind wandering detection during computerized learning in classrooms: S. Hutt et al.},
  author={Hutt, Stephen and Krasich, Kristina and Mills, Caitlin and Bosch, Nigel and White, Shelby and Brockmole, James R and D’Mello, Sidney K},
  journal={User Modeling and User-Adapted Interaction},
  volume={29},
  number={4},
  pages={821--867},
  year={2019},
  publisher={Springer}
}

@inproceedings{mota2003automated,
  title={Automated posture analysis for detecting learner's interest level},
  author={Mota, Selene and Picard, Rosalind W},
  booktitle={2003 Conference on computer vision and pattern recognition workshop},
  volume={5},
  pages={49--49},
  year={2003},
  organization={IEEE}
}

@article{ochoa2016editorial,
  title={Editorial: Augmenting learning analytics with multimodal sensory data},
  author={Ochoa, Xavier and Worsley, Marcelo},
  journal={Journal of Learning Analytics},
  volume={3},
  number={2},
  pages={213--219},
  year={2016}
}

@article{blikstein2016multimodal,
  title={Multimodal learning analytics and education data mining: Using computational technologies to measure complex learning tasks},
  author={Blikstein, Paulo and Worsley, Marcelo},
  journal={Journal of learning analytics},
  volume={3},
  number={2},
  pages={220--238},
  year={2016}
}

@article{booth2023engagement,
  title={Engagement detection and its applications in learning: a tutorial and selective review},
  author={Booth, Brandon M and Bosch, Nigel and D’Mello, Sidney K},
  journal={Proceedings of the IEEE},
  volume={111},
  number={10},
  pages={1398--1422},
  year={2023},
  publisher={IEEE}
}

@inproceedings{bosch2016detecting,
  title={Detecting student engagement: Human versus machine},
  author={Bosch, Nigel},
  booktitle={proceedings of the 2016 Conference on User Modeling Adaptation and Personalization},
  pages={317--320},
  year={2016}
}

@article{williams1949experimental,
  title={Experimental designs balanced for the estimation of residual effects of treatments},
  author={Williams, Evan James},
  journal={Australian Journal of Scientific Research Series A: Physical Sciences},
  volume={2},
  number={2},
  pages={149--168},
  year={1949},
  publisher={CSIRO Publishing}
}

@misc{mitocw,
  title={MIT OpenCourseWare},
  author={{Massachusetts Institute of Technology}},
  year={2001},
  note={\url{https://ocw.mit.edu}}
}

@inproceedings{kawsar2018esense,
  title={eSense: Open earable platform for human sensing},
  author={Kawsar, Fahim and Min, Chulhong and Mathur, Akhil and Montanari, Alessandro and Acer, Utku G{\"u}nay and Van den Broeck, Marc},
  booktitle={Proceedings of the 16th ACM Conference on Embedded Networked Sensor Systems},
  pages={371--372},
  year={2018}
}

@inproceedings{tang2025ring,
  title={?-Ring: A Smart Ring Platform for Multimodal Physiological and Behavioral Sensing},
  author={Tang, Jiankai and He, Zhe and Zhang, Mingyu and Geng, Wei and Zhou, Chengchi and Shi, Weinan and Shi, Yuanchun and Wang, Yuntao},
  booktitle={Companion of the 2025 ACM International Joint Conference on Pervasive and Ubiquitous Computing},
  pages={1271--1277},
  year={2025}
}

@article{yoon2026consensus,
  title={ConSensus: Multi-Agent Collaboration for Multimodal Sensing},
  author={Yoon, Hyungjun and Malekzadeh, Mohammad and Lee, Sung-Ju and Kawsar, Fahim and Qendro, Lorena},
  journal={arXiv preprint arXiv:2601.06453},
  year={2026}
}

@article{saha2025pulse,
  title={Pulse-ppg: An open-source field-trained ppg foundation model for wearable applications across lab and field settings},
  author={Saha, Mithun and Xu, Maxwell A and Mao, Wanting and Neupane, Sameer and Rehg, James M and Kumar, Santosh},
  journal={Proceedings of the ACM on Interactive, Mobile, Wearable and Ubiquitous Technologies},
  volume={9},
  number={3},
  pages={1--35},
  year={2025},
  publisher={ACM New York, NY, USA}
}

@article{li2024electrocardiogram,
  title={An electrocardiogram foundation model built on over 10 million recordings with external evaluation across multiple domains},
  author={Li, Jun and Aguirre, Aaron and Moura, Junior and Liu, Che and Zhong, Lanhai and Sun, Chenxi and Clifford, Gari and Westover, Brandon and Hong, Shenda},
  journal={arXiv preprint arXiv:2410.04133},
  year={2024}
}

@article{jiang2024neurolm,
  title={NeuroLM: A universal multi-task foundation model for bridging the gap between language and EEG signals},
  author={Jiang, Wei-Bang and Wang, Yansen and Lu, Bao-Liang and Li, Dongsheng},
  journal={arXiv preprint arXiv:2409.00101},
  year={2024}
}

@article{luo2024toward,
  title={Toward foundation model for multivariate wearable sensing of physiological signals},
  author={Luo, Yunfei and Chen, Yuliang and Salekin, Asif and Rahman, Tauhidur},
  journal={ACM Transactions on Computing for Healthcare},
  year={2024},
  publisher={ACM New York, NY}
}

@inproceedings{goswami2024moment,
  title={MOMENT: A Family of Open Time-series Foundation Models},
  author={Mononito Goswami and Konrad Szafer and Arjun Choudhry and Yifu Cai and Shuo Li and Artur Dubrawski},
  booktitle={International Conference on Machine Learning},
  year={2024}
}

@article{ordonez2016deep,
  title={Deep convolutional and lstm recurrent neural networks for multimodal wearable activity recognition},
  author={Ord{\'o}{\~n}ez, Francisco Javier and Roggen, Daniel},
  journal={Sensors},
  volume={16},
  number={1},
  pages={115},
  year={2016},
  publisher={MDPI}
}

@article{SEECK20172070,
title = {The standardized EEG electrode array of the IFCN},
journal = {Clinical Neurophysiology},
volume = {128},
number = {10},
pages = {2070-2077},
year = {2017},
issn = {1388-2457},
doi = {https://doi.org/10.1016/j.clinph.2017.06.254},
url = {https://www.sciencedirect.com/science/article/pii/S1388245717304832},
author = {Margitta Seeck and Laurent Koessler and Thomas Bast and Frans Leijten and Christoph Michel and Christoph Baumgartner and Bin He and Sándor Beniczky}
}

@article{jacobs1991adaptive,
  title={Adaptive mixtures of local experts},
  author={Jacobs, Robert A and Jordan, Michael I and Nowlan, Steven J and Hinton, Geoffrey E},
  journal={Neural computation},
  volume={3},
  number={1},
  pages={79--87},
  year={1991},
  publisher={MIT Press}
}

@article{arevalo2017gated,
  title={Gated multimodal units for information fusion},
  author={Arevalo, John and Solorio, Thamar and Montes-y-G{\'o}mez, Manuel and Gonz{\'a}lez, Fabio A},
  journal={arXiv preprint arXiv:1702.01992},
  year={2017}
}

@inproceedings{he2026characterizing,
  title={Characterizing Personality from Eye-Tracking: The Role of Gaze and Its Absence in Interactive Search Environments},
  author={He, Jiaman and Micheli, Marta and Spina, Damiano and McKay, Dana and Trippas, Johanne R and Kando, Noriko},
  booktitle={Proceedings of the 2026 Conference on Human Information Interaction and Retrieval},
  pages={193--203},
  year={2026}
}

@inproceedings{he2025characterising,
  title={Characterising Topic Familiarity and Query Specificity Using Eye-Tracking Data},
  author={He, Jiaman and Leng, Zikang and McKay, Dana and Trippas, Johanne R and Spina, Damiano},
  booktitle={Proceedings of the 48th International ACM SIGIR Conference on Research and Development in Information Retrieval},
  pages={2602--2606},
  year={2025}
}

@article{liu2024calibread,
  title={Calibread: Unobtrusive eye tracking calibration from natural reading behavior},
  author={Liu, Change and Yu, Chun and Wang, Xiangyang and Jiang, Jianxiao and Yang, Tiaoao and Tang, Bingda and Shi, Yingtian and Liang, Chen and Shi, Yuanchun},
  journal={Proceedings of the ACM on Interactive, Mobile, Wearable and Ubiquitous Technologies},
  volume={8},
  number={4},
  pages={1--30},
  year={2024},
  publisher={ACM New York, NY, USA}
}

@inproceedings{liu2025enhancing,
  title={Enhancing Smartphone Eye Tracking with Cursor-Based Interactive Implicit Calibration},
  author={Liu, Chang and Wang, Xiangyang and Yu, Chun and Shi, Yingtian and Wang, Chongyang and Liu, Ziqi and Liang, Chen and Shi, Yuanchun},
  booktitle={Proceedings of the 2025 CHI Conference on Human Factors in Computing Systems},
  pages={1--22},
  year={2025}
}

@article{van2024mitigating,
  title={Mitigating data quality challenges in ambulatory wrist-worn wearable monitoring through analytical and practical approaches},
  author={Van Der Donckt, Jonas and Vandenbussche, Nicolas and Van Der Donckt, Jeroen and Chen, Stephanie and Stojchevska, Marija and De Brouwer, Mathias and Steenwinckel, Bram and Paemeleire, Koen and Ongenae, Femke and Van Hoecke, Sofie},
  journal={Scientific Reports},
  volume={14},
  number={1},
  pages={17545},
  year={2024},
  publisher={Nature Publishing Group UK London}
}

@article{owen2005n,
  title={N-back working memory paradigm: A meta-analysis of normative functional neuroimaging studies},
  author={Owen, Adrian M and McMillan, Kathryn M and Laird, Angela R and Bullmore, Ed},
  journal={Human brain mapping},
  volume={25},
  number={1},
  pages={46--59},
  year={2005},
  publisher={Wiley Online Library}
}

@inproceedings{yan2022scalability,
  title={Scalability, sustainability, and ethicality of multimodal learning analytics},
  author={Yan, Lixiang and Zhao, Linxuan and Gasevic, Dragan and Martinez-Maldonado, Roberto},
  booktitle={LAK22: 12th international learning analytics and knowledge conference},
  pages={13--23},
  year={2022}
}

@article{vsvabensky2026open,
  title={Open Datasets in Learning Analytics: Trends, Challenges, and Best PRACTICE},
  author={{\v{S}}v{\'a}bensk{\`y}, Valdemar and Flanagan, Brendan and L{\'o}pez Zapata, Erwin Daniel and Shimada, Atsushi},
  journal={ACM Transactions on Knowledge Discovery from Data},
  year={2026},
  publisher={ACM New York, NY}
}

@article{skinner2009motivational,
  title={A motivational perspective on engagement and disaffection: Conceptualization and assessment of children's behavioral and emotional participation in academic activities in the classroom},
  author={Skinner, Ellen A and Kindermann, Thomas A and Furrer, Carrie J},
  journal={Educational and psychological measurement},
  volume={69},
  number={3},
  pages={493--525},
  year={2009},
  publisher={Sage Publications Sage CA: Los Angeles, CA}
}

@article{levordashka2023measuring,
  title={Measuring real-time cognitive engagement in remote audiences},
  author={Levordashka, Ana and Stanton Fraser, Dana{\"e} and Gilchrist, Iain D},
  journal={Scientific Reports},
  volume={13},
  number={1},
  pages={10516},
  year={2023},
  publisher={Nature Publishing Group UK London}
}

@article{pereira2024attention,
  author  = {Pereira, Effie J. and Ayers-Glassey, Sarah and Wammes, Jeffrey D. and Smilek, Daniel},
  title   = {Attention in Hindsight: Using Stimulated Recall to Capture Dynamic Fluctuations in Attentional Engagement},
  journal = {Behavior Research Methods},
  year    = {2024},
  volume  = {56},
  number  = {6},
  pages   = {5354--5385},
  doi     = {10.3758/s13428-023-02273-4}
}

@techreport{aczel2026environmentalAI,
  title = {Environmental Cost of AI's Energy Use: Carbon, Water and Land Footprints},
  author = {Aczel, Miriam and Chamanara, Sanaz and Matin, Mir and Farsi, Aria and Marwala, Tshilidzi and Madani, Kaveh},
  institution = {United Nations University Institute for Water, Environment and Health},
  address = {Richmond Hill, Ontario, Canada},
  year = {2026},
  doi = {10.53328/INR26RMA002},
  url = {https://unu.edu/inweh/collection/environmental-cost-of-AIs-Enrgy-Use-Carbon-water-and-land-footprints}
}

@misc{nvidiaRTXA6000Spec,
  title = {NVIDIA RTX A6000 Graphics Card Specifications},
  author = {{NVIDIA}},
  year = {2026},
  howpublished = {\url{https://www.nvidia.com/en-us/products/workstations/rtx-a6000/}},
  note = {Accessed 2026-07-23}
}

@misc{museSAthenaPrice,
  title = {Muse S Athena},
  author = {{Interaxon Inc.}},
  year = {2026},
  howpublished = {\url{https://choosemuse.com/products/muse-s-athena}},
  note = {Accessed 2026-07-23}
}

@misc{polarH10Price,
  title = {Polar H10 Heart Rate Sensor},
  author = {{Polar Electro}},
  year = {2026},
  howpublished = {\url{https://www.polar.com/us-en/sensors/h10-heart-rate-sensor}},
  note = {Accessed 2026-07-23}
}

@misc{beamEyeTrackerPrice,
  title = {Beam Eye Tracker},
  author = {{Eyeware Tech SA}},
  year = {2026},
  howpublished = {\url{https://beam.eyeware.tech/products/beam-eye-tracker}},
  note = {Accessed 2026-07-23}
}

@misc{logitechBrio100Price,
  title = {Brio 100 Full HD 1080p Webcam},
  author = {{Logitech}},
  year = {2026},
  howpublished = {\url{https://www.logitech.com/en-us/shop/p/brio-100-webcam}},
  note = {Accessed 2026-07-23}
}

@misc{microsoftBand2Price,
  title = {Microsoft Announces New Lumia Phones, Surface and Microsoft Band Devices},
  author = {{Microsoft}},
  year = {2015},
  howpublished = {\url{https://blogs.windows.com/devices/2015/10/06/microsoft-announces-new-lumia-phones-surface-and-microsoft-band-devices-taking-windows-10-experiences-to-the-next-level/}},
  note = {Accessed 2026-07-23}
}

@article{murphy2023effect,
author = {Dillon H. Murphy and Kara M. Hoover and Alan D. Castel},
title = {The effect of video playback speed on learning and mind-wandering in younger and older adults},
journal = {Memory},
volume = {31},
number = {6},
pages = {802--817},
year = {2023},
publisher = {Routledge},
doi = {10.1080/09658211.2023.2198326},
note ={PMID: 37017554},
URL = { 
        https://doi.org/10.1080/09658211.2023.2198326
},
eprint = { 
        https://doi.org/10.1080/09658211.2023.2198326}
        }

@article{gupta2016daisee,
  title={Daisee: Towards user engagement recognition in the wild},
  author={Gupta, Abhay and D'Cunha, Arjun and Awasthi, Kamal and Balasubramanian, Vineeth},
  journal={arXiv preprint arXiv:1609.01885},
  year={2016}
}

@inproceedings{kaur2018prediction,
  title={Prediction and localization of student engagement in the wild},
  author={Kaur, Amanjot and Mustafa, Aamir and Mehta, Love and Dhall, Abhinav},
  booktitle={2018 Digital Image Computing: Techniques and Applications (DICTA)},
  pages={1--8},
  year={2018},
  organization={IEEE}
}

@inproceedings{wu2024cmose,
  title={CMOSE: comprehensive multi-modality online student engagement dataset with high-quality labels},
  author={Wu, Chi-Hsuan and Liu, Shih-yang and Huang, Xijie and Wang, Xingbo and Zhang, Rong and Minciullo, Luca and Yiu, Wong Kai and Kwan, Kenny and Cheng, Kwang-Ting},
  booktitle={Proceedings of the IEEE/CVF Conference on Computer Vision and Pattern Recognition},
  pages={4636--4645},
  year={2024}
}

@article{marquez2025dipser,
  title={DIPSER: A dataset for in-person student engagement recognition in the wild},
  author={Marquez-Carpintero, Luis and Suescun-Ferrandiz, Sergio and {\'A}lvarez, Carolina Lorenzo and Fernandez-Herrero, Jorge and Viejo, Diego and Roig-Vila, Rosabel and Cazorla, Miguel},
  journal={arXiv preprint arXiv:2502.20209},
  year={2025}
}

@article{madsen2026brain,
  title={The brain, body, and behavior dataset (BBBD): Multimodal recordings during educational videos},
  author={Madsen, Jens and Kuppa, Nikhil and Parra, Lucas C},
  journal={Scientific Data},
  year={2026},
  publisher={Nature Publishing Group UK London}
}
